\documentclass[12pt,a4paper]{article}

\usepackage[utf8]{inputenc}
\usepackage[english]{babel}
\usepackage{amsmath}
\usepackage{amsfonts}
\usepackage{dsfont}
\usepackage{graphicx}
\usepackage{indentfirst}
\usepackage{subfigure}
\usepackage{float}
\usepackage{xcolor}
\usepackage{setspace}
\usepackage{url}
\usepackage[authoryear]{natbib}

\DeclareMathOperator*{\arginf}{arg\,inf}

\newtheorem{theorem}{Theorem}[section]
\newtheorem{lemma}{Lemma}[section]
\newtheorem{remark}{Remark}[section]

\newtheorem{example}{Example}[section]

\definecolor{orange}{rgb}{1,0.5,0}
\definecolor{electricpurple}{rgb}{0.75, 0.0, 1.0}
\definecolor{green(ryb)}{rgb}{0.4, 0.69, 0.2}

\begin{document}

\title{Optimal discrimination designs for semi-parametric models }

\author{
{\small Holger Dette } \\
{\small Ruhr-Universit\"at Bochum } \\
{\small Fakult\"at f\"ur Mathematik } \\
{\small 44780 Bochum, Germany } \\
{\small e-mail: holger.dette@ruhr-uni-bochum.de }\\
\and
{\small Roman Guchenko, Viatcheslav Melas } \\
{\small  St. Petersburg State University} \\
{\small  Faculty of Mathematics and Mechanics} \\
{\small  St. Petersburg, Russia} \\
{\small e-mail: romanguchenko@ya.ru,  vbmelas@post.ru} \\
\and
{\small Weng Kee Wong } \\
{\small Department of Biostatistics } \\
{\small UCLA Fielding School of Public Health } \\
{\small  Los Angeles, CA 90095-1772 } \\
{\small e-mail: wkwong@ucla.edu }
}

\maketitle
\begin{abstract}
Much of the work in the literature on optimal discrimination designs assumes that the models of interest are fully specified, apart from  unknown parameters in some models.  Recent work  allows errors in the models to be non-normally distributed but still requires the specification  of the mean structures.
This research  is motivated by the interesting work of \cite{Otsu2008}  to discriminate among semi-parametric models by  generalizing the KL-optimality criterion proposed by \cite{loptomtra2007} and \cite{tomlop2010}.   In our work we provide further important insights in this interesting optimality criterion. In particular, we propose a practical strategy for finding optimal discrimination designs among semi-parametric models that can also be verified using an equivalence theorem. In addition, we study properties of such optimal   designs and  identify important cases where  the proposed semi-parametric optimal discrimination designs coincide with the celebrated $T$-optimal designs.
\end{abstract}

Keywords and Phrases:  continuous design, discrimination design,   equivalence theorem, semi-parametric model, $T$-optimality, variational calculus.

\section{Introduction}
\label{introduction}
\def\theequation{1.\arabic{equation}}
\setcounter{equation}{0}

Finding optimal discrimination designs dates back to 1970's and probably earlier.  Early references are
\cite{stigler1971} who considered nested models and   \cite{atkfed1975a,atkfed1975b}, where they  proposed  $T$-optimal designs to  discriminate among models when errors are normally distributed. The first of the two last named papers concerns discrimination between two models and the second paper generalized the problem to multiple models.  $T$-optimality   for  discriminating between  parametric models assumes that  we have a known null model and we wish to test  whether a rival model with unknown parameter holds.
A likelihood ratio test  is then used to discriminate between the two rival models  and  under  local alternatives,  the   non-centrality parameter of the chi-square distribution of the
   test statistic, which  contains the unknown parameters from the alternative model,  is proportional to the $T$-optimality criterion \citep{wiens2009}.  Since a larger non-centrality parameter provides a more powerful test, the $T$-optimal design  strategy is to find a design that maximizes the minimum value of the non-centrality parameter, where the minimum is taken over all possible values of the parameters in the model under the alternative hypothesis.  Consequently,  $T$-optimality criterion   is a type of maximin optimality criterion  and designs that maximize the minimum of the non-centrality parameter are called $T$-optimal designs.

The two seminal  papers of  \cite{atkfed1975a,atkfed1975b}  contain   illustrative examples.  Because the $T$-optimality criterion is not differentiable,  calculations  to determine the optimal design can be involved  even for relatively simple models; see for example, \cite{detmelshp2012} and \cite{detmelshp2016} for some  explicit non-trivial results.
For the same reason,  the construction
of efficient algorithms for finding $T$-optimal designs is a challenging problem.
  Some recent progress in this direction includes
\cite{BraessDette2013,detmelguc2015,detmelguc2016} and \cite{martinmartin}.  An interactive software tool for finding optimal discrimination designs is available in \cite{stegmaier}.

There  is notable  progress in  tackling optimal discrimination design problems on various fronts; the progress appeared periodic  a few decades ago and seems to have picked up in recent years. Advances include alternative problem formulations and broader applications of such designs in cognitive science \citep{Covagnaro2010}, psychology \citep{Myung2009} and chemical engineering \citep{Alberton2011}, to name a few. Different and more flexible optimality criteria have also been proposed.  In particular, advances in the construction of optimal discrimination designs include the following: (i) the frequently criticized unrealistic assumption in the $T$-optimality criterion that requires a known model in the null hypothesis is now removed \citep{jamsen:2013}; (ii) the class of models of interest now includes generalized linear models, where outcomes can be discrete \citep{wawoecle08}, (iii) optimal designs for discriminating multivariate models \citep{Yanagisawa1990,ucibog2005}; (iv) Bayesian optimal designs for model discrimination \citep{felsenstein,tomlop2010,detmelguc2015} , (v) dual-objective optimal designs for model discrimination \citep{atkbogbog1998,ng2004,atkinson2008b,Alberton2011,monsef}, (vi) discriminating models with correlated errors \citep{jesus}, (vii) adaptive designs for model discrimination \citep{Myung2009, donckels} and (viii) optimal discrimination designs for dynamic models \citep{ucibog2005}.
Other important references that describe alternative approaches and properties of optimal discrimination designs are \cite{loptomtra2007,dettit2009,detmelguc2015}, among others.

All references cited so far require a  parametric specification of the conditional distribution of the response, which
raises some robustness issues in the application of the $T$-optimality criterion for constructing optimal
discrimination  designs.
 Robustness properties of optimal discrimination designs
with respect  to various model assumptions have been considered by  \cite{wiens2009,ghosh2013} and \cite{DetteMelasShpilev2013}.
\cite{Otsu2008}  proposed a new optimality criterion for   discriminating between models, which is similar in spirit to
the classical $T$-optimality criterion and its extensions except that it does not
require an exact specification of the conditional distribution. Optimal discrimination  designs were found  using the duality relationships in entropy-like minimization problems \citep{BorweinLewis1991} and the resulting optimal designs  are called semi-parametric optimal discrimination  designs.

The present paper provides a more careful analysis of the novel approach proposed by \cite{Otsu2008}. We  propose a  more practical strategy for finding semi-parametric
optimal discrimination designs and derive several important properties of such optimal designs.
In Section \ref{sec2}, we  define semi-parametric optimal discrimination  designs
  and derive several auxiliary results that substantially simplify the calculation of
optimal   designs.  In Section \ref{sec3}, we  provide  equivalence theorems
to  characterize   semi-parametric optimal discrimination  designs and demonstrate that all designs derived by  \cite{Otsu2008} are in fact not optimal. Section \ref{sec4} describes the relation between  semi-parametric optimal discrimination  designs and  optimal discrimination designs  under some  criteria discussed in the literature. In particular, we identify
cases, where the semi-parametric optimal discrimination  designs  proposed  by \cite{Otsu2008} coincide with the $T$-optimal discrimination  designs and KL-optimal designs  introduced  by \cite{atkfed1975a,atkfed1975b}  and  \cite{loptomtra2007}, respectively.  Section \ref{sec5} presents numerical results and all technical details are deferred to the appendix in Section \ref{sec6}.\\

\section{Semi-parametric discrimination designs}
\label{sec2}
\def\theequation{2.\arabic{equation}}
\setcounter{equation}{0}

We model the response variable $Y$ as a function of a vector of explanatory variables $x$ defined on a given compact design space $\mathcal{X}$.  Suppose the density of $Y$ with respect to
the  Lebesgue measure is $f(y,x)$,    and we have resources to take a fixed number of observations, say $n$, for the study.   Following \cite{kiefer1974}, we focus on approximate
designs  which are essentially probability measures defined on $\mathcal{X}$. If an approximate design  has $k$ support points, say
  $ x_1,\ldots, x_k $,   and the corresponding weights are $\omega_1,\ldots, \omega_k$, then approximately $n \omega_i$ observations are taken at $x_i, i=1,\ldots,k$. In practice, a rounding procedure is applied to every $n \omega_i$ so that they are positive  integers $n_{i} $ ($i=1,\ldots,k)$ and $\sum_{i=1}^kn_i=n$.  The experimenter then takes $n_i$
independent  observations with $Y_{i,1}, \ldots , Y_{i,n_i}$ at $x_i$ which are assumed to  have a density $f(y,x_i)$ with respect to the Lebesgue measure, $i=1,\ldots , k$.

To construct efficient designs for discriminating between  two competing models for $f(y,x)$,  \cite{loptomtra2007}  assumed parametric densities, say     $f_j(y,x,\theta_j)$, where the  parameter $\theta_j$ varies in a compact parameter space, say 
 $\Theta_j,$ $ j=1,2$.  To fix ideas, we ignore nuisance parameters which may be present in the models. The  Kullback-Leibler distance between two densities $f_{1}$ and $f_{2}$ is
  \begin{equation} \label{KL1}
 I_{ 1,2} (x, f_{ 1}, f_{ 2}, \theta_{ 1}, \theta_{2}) =  \int f_{1} (y, x, \theta_{1}) \log
 \frac {f_{1} (y,x, \theta_{1})}{f_{2}(y,x,\theta_{2})} dy
 \end{equation}
 and it measures the discrepancy between the densities.
  \cite{loptomtra2007} assumed that the model  $f_{1}$
 is  the ``true'' model with a fixed parameter vector $\overline \theta_{1}$ and
  call a design local  {\it KL-optimal discriminating design for the models $f_{1}$ and $f_{2}$} if
  it  maximizes the criterion
 \begin{equation} \label{KL2}
\mathrm{KL}_{1,2} (\xi, \overline \theta_{1}) = \inf_{\theta_{2}\in \Theta_2}
\int_{\mathcal{X}} I_{1,2} (x, f_{1},f_{2} ,  {\overline{\theta}_{1}}, \theta_{2}) \xi (dx).
 \end{equation}

\cite{Otsu2008} considered a novel setup and proposed a design criterion for  discriminating a parametric model defined by its density and another nonparametric model.  More precisely, this author considered
the design problem for testing the hypothesis
 \begin{equation} \label{hyp}
 H_0:  \eta(x) = \eta_1(x,\theta_1) ~~\mbox{versus} ~~ H_1:   \eta(x) =\eta_2(x,\theta_2)
\end{equation}
where
 $$
 \eta_j (x, \theta_j) =  \int  y f_j (y,x,\theta_j)  dy, \; j = 1,2,
 $$
is the conditional mean of the density $f_j(y,x,\theta_j)$ with support set
\begin{equation} \label{supp}
{\cal S}_{f_j, \theta_j , x}  =  \big\{ y   ~|~ f_j(y,x,\theta_j) > 0 \big\} ~,~~
j=1,2.
\end{equation}
  The setup   is therefore more general than that in
 \cite{loptomtra2007}, who assumed that $f_1$ and $f_2$ are known and   one of the  parametric models is fully specified.
  To be more specific, let $f_1(y,x, \overline\theta_1)$ be a parametric density with a fixed parameter $\overline\theta_1$ and define

 \begin{align}
 \label{F2}
\mathcal{F}_{2 ,x,\theta_2} = \Big \{ f_2 : \int f_2(y,x,\theta_2)\, dy = 1, \; \int y f_2(y,x,\theta_2) dy = \eta_2(x,\theta_2), ~\mathcal{S}_{f_2, \theta_2 , x}  = \mathcal{S}_{f_1,
\overline{\theta}_1 ,  x}  \Big \},
\end{align}
which is the class of all conditional densities (at the point $x$) with parameter $\theta_2$ and  conditional mean $ \eta_2(x,\theta_2)$.
Consider the set obtained from $\mathcal{F}_{2 ,x,\theta_2}$ by letting the ranges of $x$ and $\theta_2$ vary over all
their possible values:
\begin{align*}
\mathcal{F}_{2 }  = \big\{ \mathcal{F}_{2 ,x,\theta_2} ~|~x \in {\cal X};~\theta_2 \in \Theta_2 \big\}
\end{align*}
and call a design $\xi^*$ {\it semi-parametric
optimal  design for discriminating between   the model  $f_1(y,x,  \overline\theta_1)$ and models in the class $\mathcal{F}_2$} if it maximizes the optimality criterion
\begin{align} \label{f1F2}
K_{(a)}(\xi,{\overline\theta_1}) = \inf_{\theta_2 \in \Theta_2} \int_{X} \inf_{f_2 \in \mathcal{F}_{2, x,\theta_2} }
I_{1,2} (x, f_1,f_2 ,  { \overline{\theta}_1}, \theta_2)
 \, \xi(dx)
\end{align}
among all approximate designs defined on the design space ${\cal X} $. Note that this criterion is a local
optimality criterion in the sense of \cite{chernoff1953} as it depends on the parameter $\overline\theta_1$.  The subscript $(a)$ denotes the first   design criterion  with the subscript $(b)$ for the second design criterion  below. \\
Similarly,  we may fix the conditional density $f_2(y,x,\theta_2)$ and
define
\begin{align} \label{F1}
\mathcal{F}_{1, x,\overline\theta_1}  &= \Big \{ f_1 : \int f_1(y,x,{  \overline\theta_1})\, dy = 1,  \int y f_1(y,x,{  \overline\theta_1}) {  dy} = \eta_1(x,{  \overline\theta_1}),
~ \mathcal{S}_{f_1,
\overline{\theta}_1 ,  x}  = \mathcal{S}_{f_2, \theta_2 , x}
 \Big\},
\end{align}
which is the class of all conditional densities with parameter $\overline{\theta}_1$ and conditional mean $ \eta_1(x,\overline{\theta}_1)$.  For fixed $\overline\theta_1$, let
\begin{align*}
\mathcal{F}_{1 }  &= \big\{ \mathcal{F}_{1 ,x,\overline\theta_1} ~|~x \in {\cal X} \big\}
\end{align*}
and  we call   a design $\xi$ locally {\it semi-parametric
optimal  design  for discriminating between   the model $f_2(y,x, \theta_2)$ and the class $\mathcal{F}_{1}$} if it maximizes the optimality criterion
\begin{align}  \label{f2F1}
K_{(b)}(\xi, {\overline\theta_1}) = \inf_{\theta_2 \in \Theta_2} \int_{X} \inf_{f_1 \in \mathcal{F}_{1 ,x,\overline\theta_1}} I_{1,2} (x, f_1,f_2 ,  { \overline\theta_1}, \theta_2) \, \xi(dx),
\end{align}
among all approximate designs defined  on the design space ${\cal X} $.

 \begin{remark}{\rm  We note that the above approach can be similarly applied when the response variable $Y$ is discrete and it has one of two possible discrete probability measures that depends on $x$ and $\overline\theta_1$.   Let these two  competing measures with the same support be
\begin{align*}
P(x,\overline\theta_1) =
\begin{bmatrix}
	y_1(x,\overline\theta_1) & \dots & y_s(x,\overline\theta_1)\\
	p_1(x,\overline\theta_1) & \dots & p_s(x,\overline\theta_1)	
\end{bmatrix} \;~\text{and}~
Q(x,\theta_2) =
\begin{bmatrix}
	y_1(x,\overline\theta_1) & \dots & y_s(x,\overline\theta_1)\\
	q_1(x,\theta_2) & \dots & q_s(x,\theta_2)	
\end{bmatrix}.
\end{align*}
If we let $y_i = y_i(x,\overline\theta_1)$, let $p_i = p_i(x,\overline\theta_1)$ and let $q_i = q_i(x,\theta_2)$ for simplicity, we have $\sum_{i=1}^s p_i = 1=\sum_{i=1}^s q_i$ and $\sum_{i=1}^s y_i p_i = \eta_1(x,\overline\theta_1)$.  Define two sets
\begin{align*}
Q_{x,\theta_2} = \Big\{ q : \sum_{i=1}^s q_i = 1, \; \sum_{i=1}^s y_i q_i = \eta_2, \;  0 < q_i \leq 1 \Big\}\ \text{and}~ Q = \Big\{ Q_{x,\theta_2} ~|~x \in {\cal X};~\theta_2 \in \Theta_2 \Big\}.
\end{align*}
 To find a semi-parametric optimal design $\xi^*$ for discriminating between the model $P(x, \overline\theta_1)$ and the class $Q$ we may use the discrete version of the criterion~\eqref{f1F2}, which is
\begin{align*}
\tilde K_{(a)}(\xi,{\overline\theta_1}) = \inf_{\theta_2 \in \Theta_2} \int_{X} \inf_{q \in Q_{x,\theta_2} }
\sum_{i = 1}^s p_i(x,\overline\theta_1) \log{\frac{p_i(x,\overline\theta_1)}
{q_i(x,\theta_2)}}
 \, \xi(dx).
\end{align*}

We then write down the discrete version of~\eqref{f2F1} in a similar manner and proceed, omitting details for space consideration.}
\end{remark}

In what is to follow, we assume for simplicity that $f_1(y,x,\theta_1)$, $f_2(y,x,\theta_2)$, $\eta_1(x,\theta_1)$, $\eta_2(x,\theta_2)$ are differentiable with respect to $y$, $x$, $\theta_1$ and $\theta_2$, even though these assumptions can be weakened a bit for what we plan to do here.  \cite{Otsu2008} derived an explicit form  for the two criteria. For  the criterion \eqref{f1F2},  he obtained
\begin{align*}
K_{(a)}(\xi,{ \overline\theta_1}) = \inf_{\theta_2 \in \Theta_2} \int_{X} \left\{ \mu + 1 + \int \log \left\{ -\mu - \lambda (y - \eta_2(x,\theta_2)) \right\} f_1(y,x,{ \overline\theta_1}) dy \right\} \xi(dx),
\end{align*}
where the constants $\lambda$ and $\mu$  depend on $x$ , $\overline\theta_1$ and   $\theta_2$ and  are roots of  the system of equations
\begin{align*}
- \int \frac{f_1(y,x,{ \overline\theta_1})}{\mu + \lambda (y - \eta_2(x,\theta_2))} dy = 1, \;
\int \frac{(y - \eta_2(x,\theta_2))f_1(y,x,{ \overline\theta_1})}{\mu + \lambda (y - \eta_2(x,\theta_2))} dy = 0
\end{align*}
that satisfy the constraint
\begin{align*}
\mu + \lambda (y - \eta_2(x,\theta_2)) < 0  ,\; \mbox{~for all ~}  y \in  {\cal S}_{f_1, \overline\theta_1, x}.
\end{align*}
 A similar result can be obtained  for  the criterion \eqref{f2F1}  using somewhat similar arguments and  they are omitted for space consideration.
 In either case, the desired values for $\mu$ and $\lambda$ have to be found numerically. \cite{Otsu2008} did not provide an effective numerical procedure for finding  solutions in the two dimensional space that solve the system of equations and satisfy the constraint. In what is to follow, we show that the inner optimization problems in~\eqref{f1F2} and~\eqref{f2F1} can be reduced to solving a single equation, which is much faster and more reliable.
  For this purpose we derive simpler expressions for the criteria \eqref{f1F2}  and  \eqref{f2F1} that facilitate the computation of the semi-parametric optimal discriminating designs.  One of our key results is the following.

\begin{theorem}
\label{prop1} ~~\\
(a)  Assume that for each $x \in \mathcal{X}$ the support of the conditional  density $f_1(y,x,\overline\theta_1)$ is an interval,  i.e. $
  {\cal S}_{f_1, \overline\theta_1, x}
=
[y_{x,\min}, y_{x,\max}] $, such that $y_{x, \min} < \eta_2(x, \theta_2) < y_{x, \max}$ for all $\theta_2 \in \Theta_2$.  Assume further that for all $x \in \mathcal{X}$ and for all $\theta_2 \in \Theta_2$ there exists a unique non-zero solution $\overline{\lambda}(x, \overline\theta_1, \theta_2) $ of the equation
\begin{align} \label{eq:lambda}
\int \frac{f_1(y, x, \overline\theta_1)}{1 + \lambda(y - \eta_2(x,\theta_2))} dy = 1,
\end{align}
that satisfies
\begin{align}
\label{eq:lambda_bounds}
-\frac{1}{y_{x, \max} - \eta_2{(x, \theta_2)}} < \overline{\lambda}(x, \overline\theta_1, \theta_2) < -\frac{1}{y_{x, \min} - \eta_2 (x,\theta_2)}.
\end{align}
The criterion \eqref{f1F2} then takes the form
\begin{align}
\label{f1F2symp}
K_{(a)}(\xi, \overline{\theta}_1)  & = \inf_{\theta_2 \in \Theta_2} \int_\mathcal{X} \int \log\left\{1 + \overline{\lambda}(x, \overline\theta_1, \theta_2) (y - \eta_2(x,\theta_2)) \right\} f_1(y, x, \overline\theta_1) dy \xi(dx)
\end{align}
and the ``optimal'' density $f_2^*$ in  \eqref{f1F2} is given by
\begin{align}
\label{f1F2opt}
f_2^*(y, x, \theta_2) = \frac{f_1(y, x, \overline \theta_1)}{1 + \overline{\lambda}(x, \overline\theta_1, \theta_2) (y - \eta_2(x,\theta_2))}.
\end{align}
(b) Assume that the integrals
\begin{align*}
 \int f_2(y, x, \theta_2) \exp(-\lambda y) dy  ~\mbox{ and }  ~\int y f_2(y, x, \theta_2) \exp(-\lambda y) dy
\end{align*}
 exist for all $x \in \mathcal{X}$ and for all  $\lambda$.  The
 criterion \eqref{f2F1} takes the form
\begin{align}
\label{f2F1symp}
K_{(b)}(\xi) = \inf_{\theta_2 \in \Theta_2} \int_{\mathcal{X}} \int \left[\log\left\{\mu^\prime(\overline{\lambda}_x)\right\} - \overline{\lambda}_x y \right] f_2(y, x, \theta_2) \exp(- \overline{\lambda}_x y) \mu^\prime(\overline{\lambda}_x) dy \xi(dx)
\end{align}
where
\begin{align}
\label{f2mu}
\mu^{\prime}(\lambda) = \mu^{\prime}(\lambda, x, \theta_2) = \frac{1}{\int f_2(y, x, \theta_2) \exp(-\lambda y) dy}
\end{align}
and $\overline{\lambda}_x = \overline{\lambda}(x, \overline\theta_1, \theta_2)$ is the nonzero root of the equation
\begin{align}\label{eq:lambda2}
\frac{\int y f_2(y, x, \theta_2) \exp(-\lambda y) dy}{\int f_2(y, x, \theta_2) \exp(-\lambda y) dy} = \eta_1(x, \overline\theta_1).
\end{align}
Moreover, the ``optimal'' density $f_1^*$ in  \eqref{f2F1} is given by
\begin{align} \label{f2F1opt}
f_1^*(y, x, \overline\theta_1) = \frac{f_2(y,x,\theta_2) \exp(-\overline{\lambda}(x, {\overline\theta_1}, \theta_2) y)}{\int f_2(y,x,\theta_2) \exp(-\overline{\lambda}(x, \overline\theta_1, \theta_2) y) dy} .
\end{align}
\end{theorem}

\medskip

The main implication of Theorem \ref{prop1} is that we first solve equations  \eqref{eq:lambda}   and \eqref{eq:lambda2} numerically for $\lambda$.
For the solution of  \eqref{eq:lambda}, it is natural to presume that $\lambda < 0$  if $\eta_1(x,\overline\theta_1) < \eta_2(x,\theta_2)$.  This is because, in this case, the function
 $1/[1+\lambda (y - \eta_2(x,\theta_2))]$ is increasing whenever $y \in    {\cal S}_{f_1, \overline\theta_1, x}
 $, allowing us to shift the average of the function $f_1(y,x,\overline\theta_1)/[1+\lambda (y - \eta_2(x,\theta_2))]$ to the right. Similarly, if $\eta_1(x,\overline\theta_1) > \eta_2(x,\theta_2)$, we search for $\lambda > 0$. We state this intuitive consideration formally in the following lemma, whose proof is deferred to the appendix. 

\begin{lemma}
\label{lemma:lambda_sign}
Assume that $v_2^2(x,\theta_2) = \int (y - \eta_2(x,\theta_2))^2  f_2(y, x, \theta_2) dy $ exists and is positive.
If  $\overline{\lambda}$ is a  solution of the equation~(\ref{eq:lambda})
 and  satisfies~(\ref{eq:lambda_bounds}), then
 $\overline{\lambda} $ has the same sign as the difference   $\eta_1(x,\theta_1)  - \eta_2(x,\theta_2)$.
\end{lemma}

\medskip

\begin{figure}[H]
\centering
\subfigure{%
\includegraphics[width=50mm]{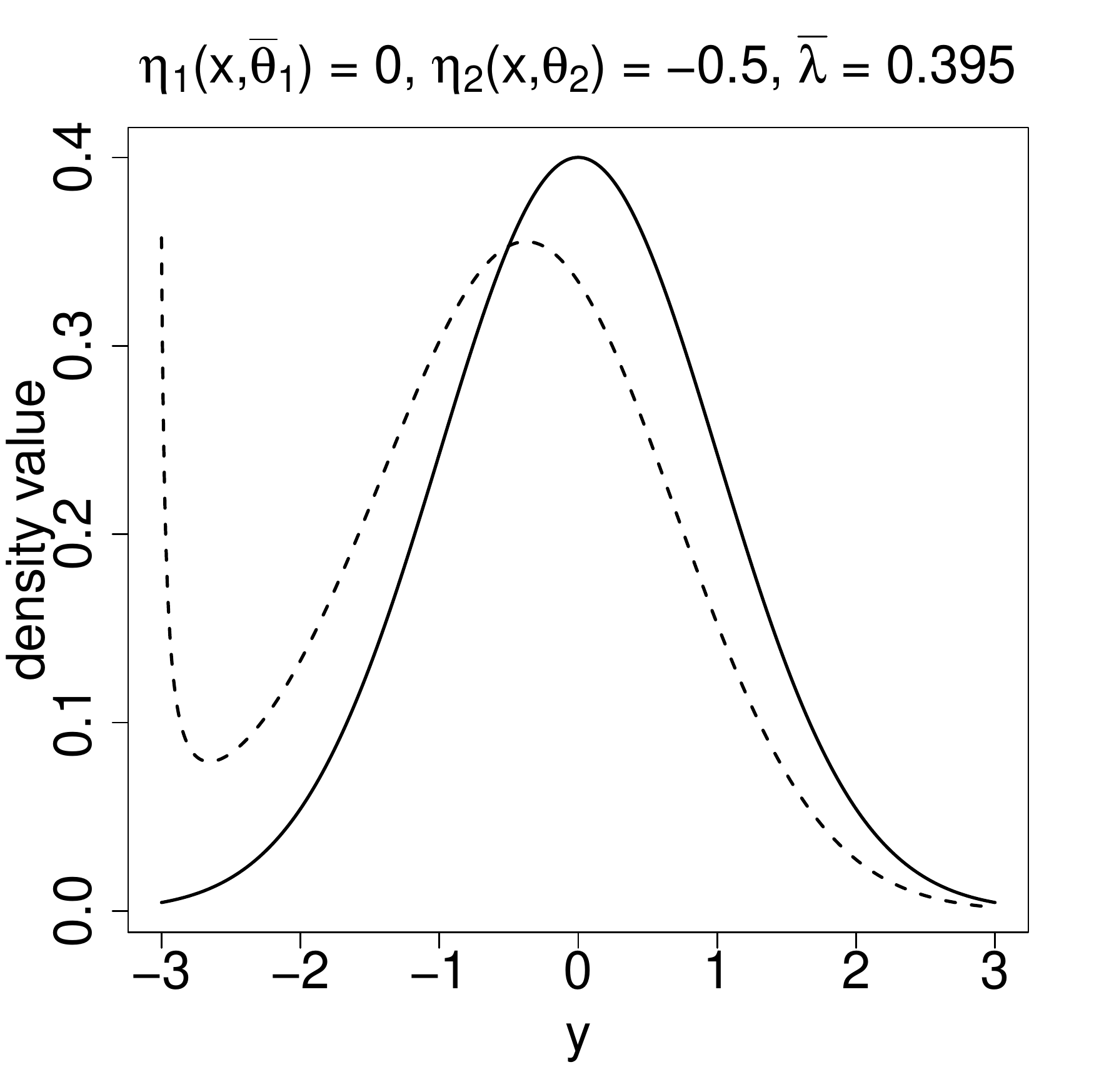}
\label{fig:subfigure1}}
\subfigure{%
\includegraphics[width=50mm]{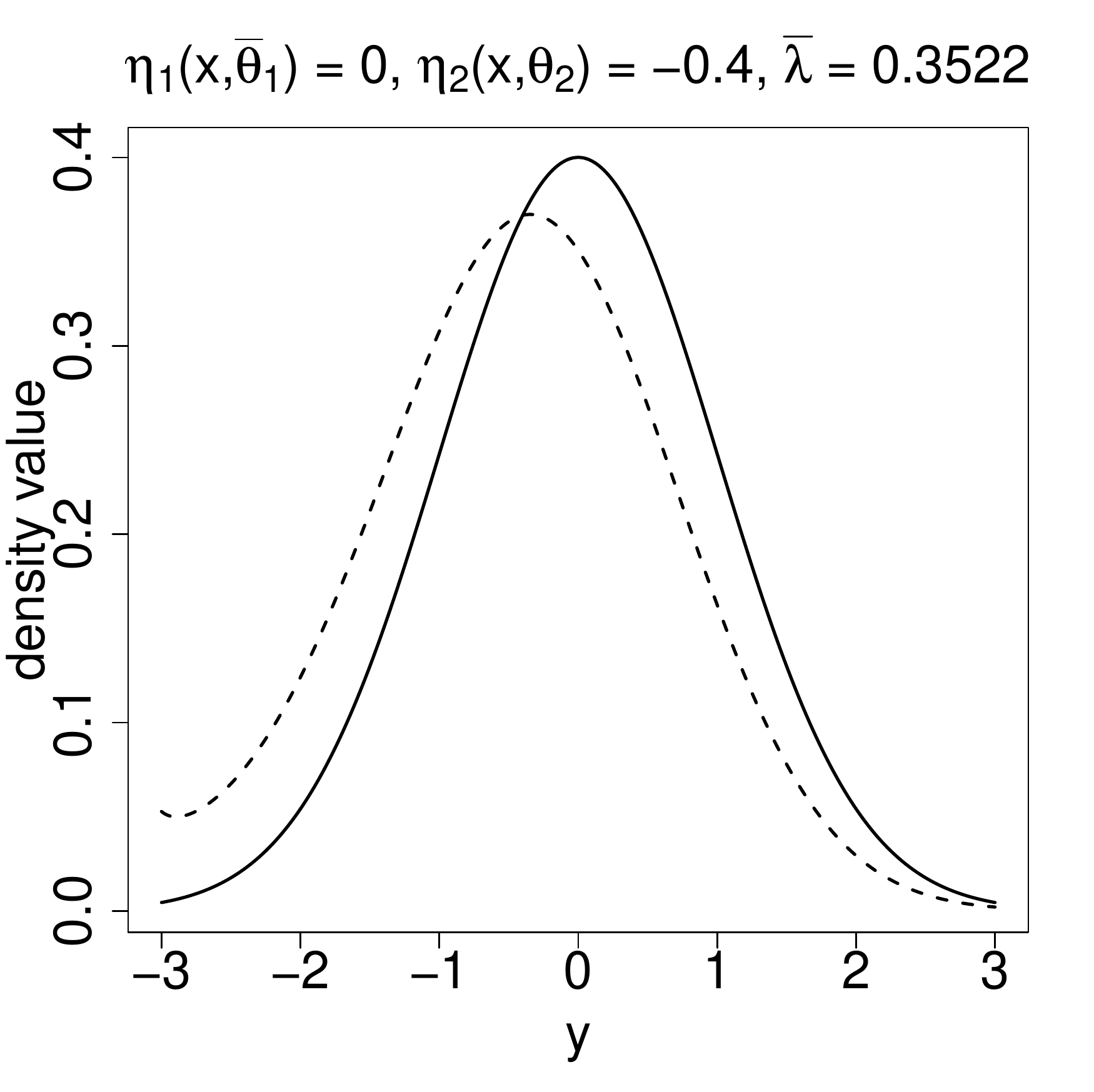}
\label{fig:subfigure2}}
\subfigure{%
\includegraphics[width=50mm]{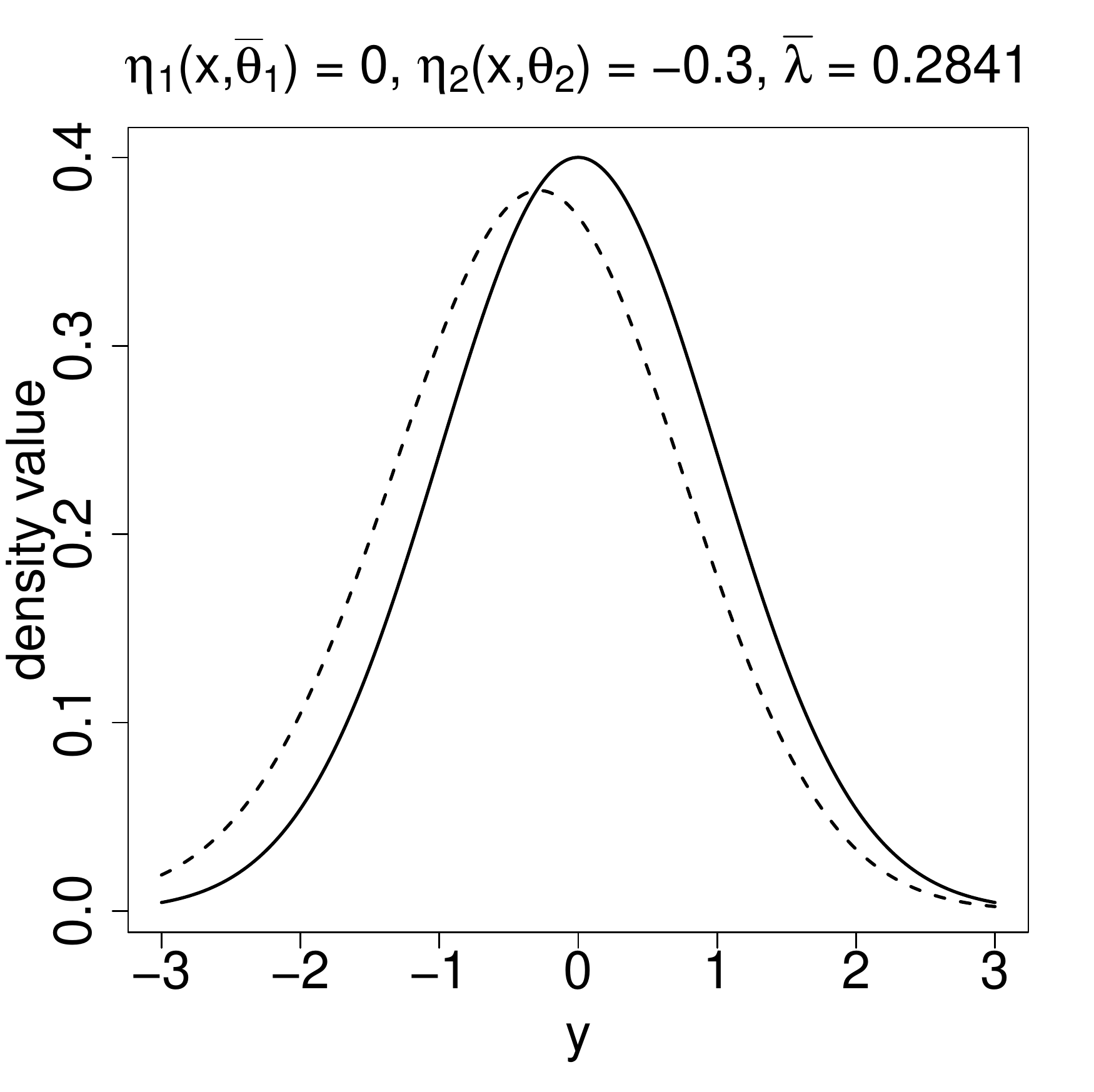}
\label{fig:subfigure3}}
\subfigure{%
\includegraphics[width=50mm]{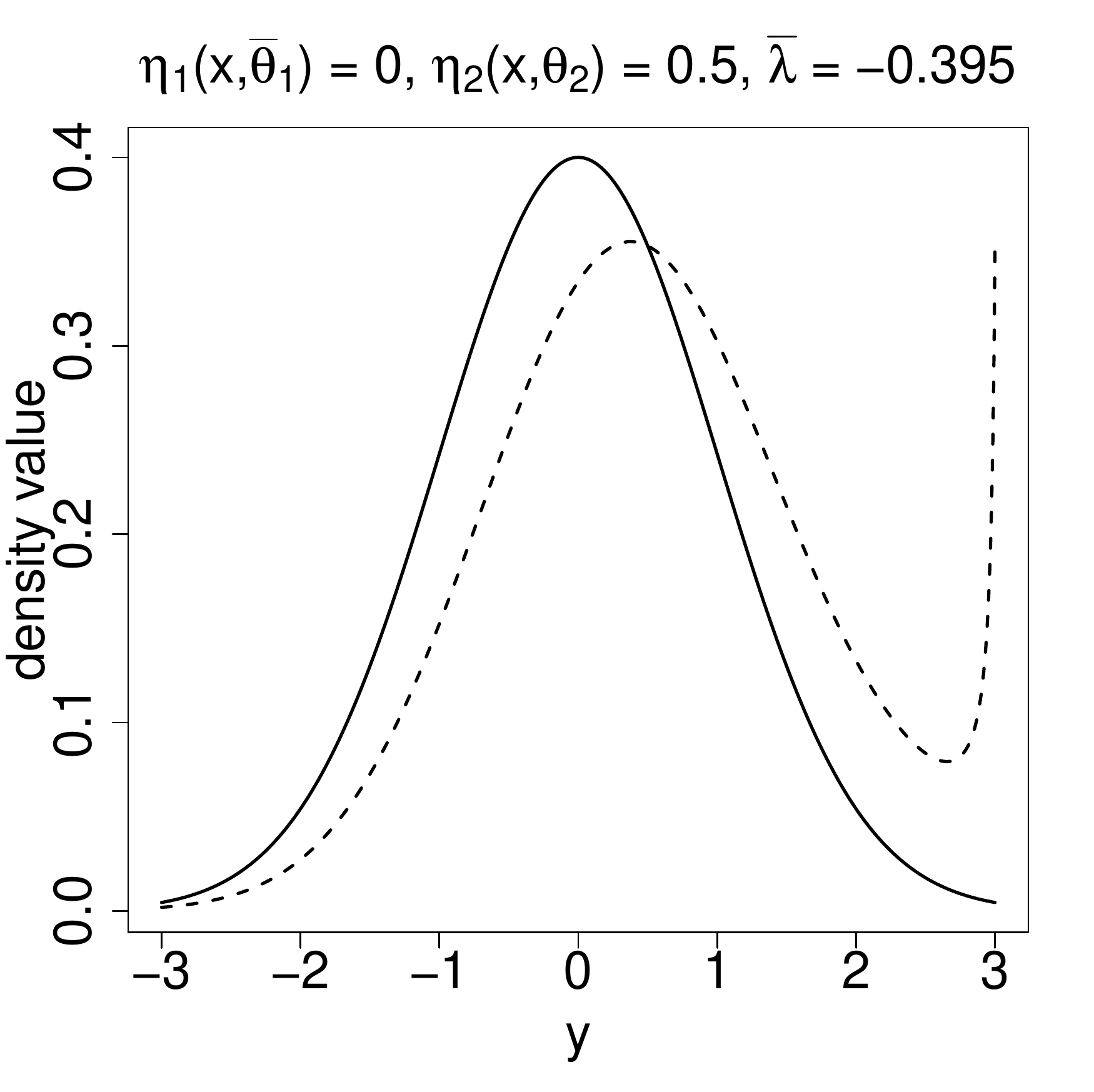}
\label{fig:subfigure4}}
\subfigure{%
\includegraphics[width=50mm]{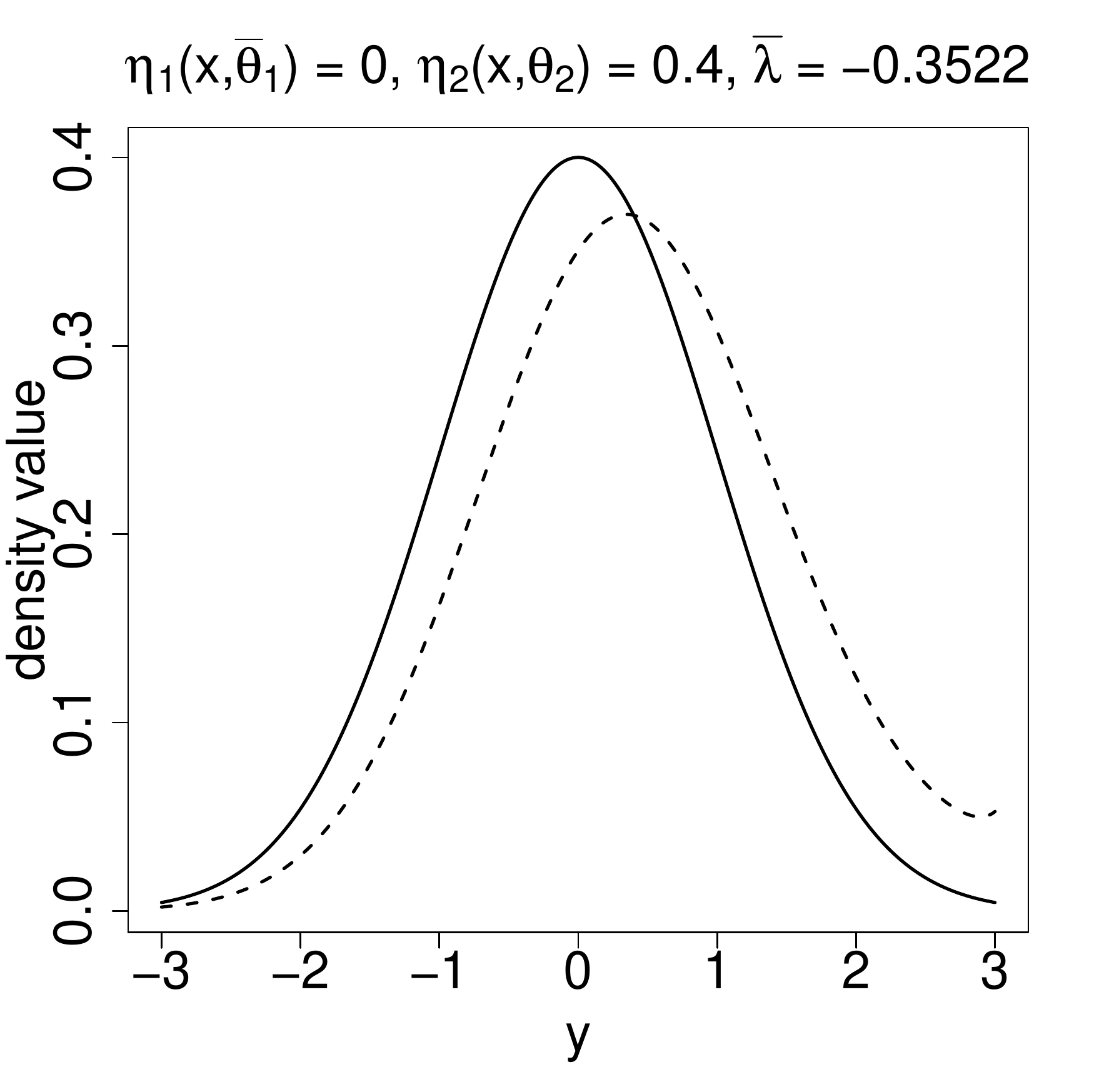}
\label{fig:subfigure5}}
\subfigure{%
\includegraphics[width=50mm]{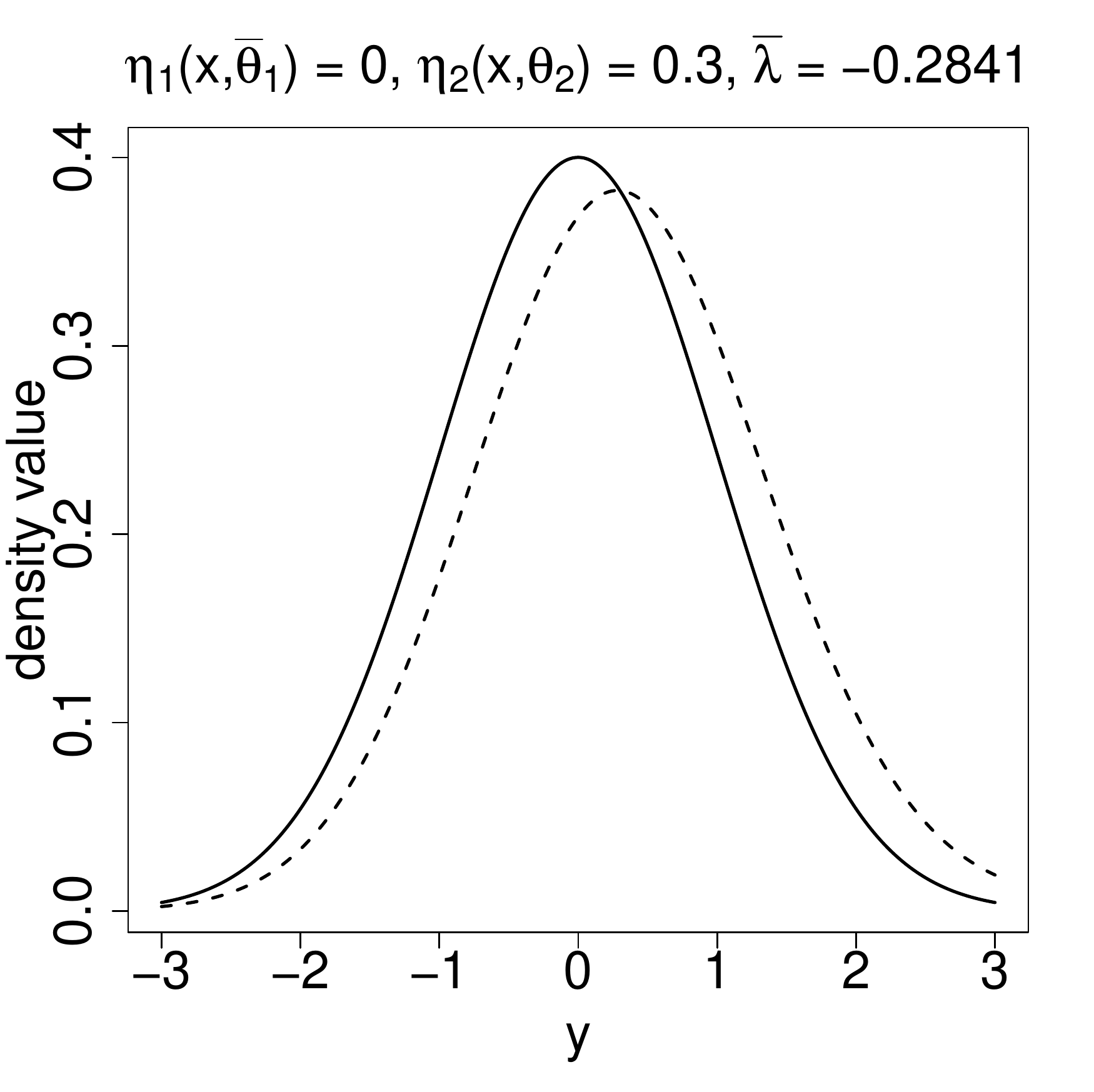}
\label{fig:subfigure6}}
\label{fig:trunc_norm}
\caption{\it The density $f_1$ (solid line) of a truncated standard normal distribution on the interval $[-3,3]$
 and the solution $f_2^*$  in \eqref{f1F2opt} for various choices of the mean function, that is  $\eta_2 (x, \theta_2)= -0.5, -0.4, -0.3, +0.5, +0.4, +0.3$.}
\end{figure}

\begin{example} \label{exam1}
{\rm
Let $f_1(y,x,\overline\theta_1)$ be the truncated normal density $\mathcal{N}(\eta(x,\overline\theta_1), 1)$ on the interval $[-3+\eta_1(x,\overline\theta_1), 3 + \eta_1(x,\overline\theta_1)]$. This density is a function of $\eta_1(x,\overline\theta_1)$
and it follows from
  \eqref{f1F2opt} that the optimal density $f_2^*(y,x,\theta_2)$ is a function of $\eta_1(x, \overline{\theta}_1)$ and $\eta_2(x,\theta_2)$  in this case. Figure~\ref{fig:trunc_norm} shows plots of the function $f_2^*$ for $\eta_1(x,\overline{\theta}_1) \equiv 0$ and different values of $\eta_2(x,\theta_2)$ on the interval $[-3,3]$.  On the top of each figure the value $\overline\lambda$ shows the solution of equation \eqref{eq:lambda}.
}
\end{example}

\section{Equivalence theorems}
\label{sec3}\def\theequation{3.\arabic{equation}}
\setcounter{equation}{0}
In our setup, the  variables we wish to optimize in the design problem are the number of  support points, the locations of the support points and the weights at each of the support point.  Because  the number of support points in the optimal design   is not known in advance, the dimension of the optimization is unknown at the onset.  A common strategy is to confine the search for the   optimal  design to a smaller class of designs, say, with a fixed number point designs.  This may simplify the search and result in a closed-form description of the optimal design; however, such an approach may not produce a design that is optimal among all designs on $\mathcal{X}$.

When the design problem is formulated as a convex or concave optimization problem, it is possible to derive an equivalence theorem to verify if a given design is optimal among all designs on the given design space.  Each convex or concave functional has a unique equivalence theorem but they all have same form with an inequality bounded above by $0$ with equality at the support points of the optimal design.  Frequently, the function on the left hand side of the inequality in the equivalence theorem is called the sensitivity function and if the dimension of the design space is small, optimality of a design can be directly verified by plotting the sensitivity function over the design space and examining the resulting plot.  Design monographs, such as \cite{pazman1986,silv80,pukelsheim2006} present equivalence theorems for various convex criteria.   When   the design criterion is concave (or convex) and not differentiable, the equivalence theorems  involve subgradients
and consequently, the optimal designs are more difficult to find and confirm via an equivalence theorem.
 However, they still have the same form as just discussed, see \cite{wong92,dette1997} and others, for example.  For $T$-optimality and its other variations, equivalence theorems are available in
\cite{loptomtra2007,dettit2009} among others.
\smallskip

Our proposed optimality  criterion is a concave functional on the space of all approximate designs  and so it is possible to derive an equivalence theorem for confirming the optimality of a design as a  semi-parametric optimal discrimination design. The next theorem presents the equivalence theorems for checking the optimality of a design under the two new criteria. A major difference here  compared to the ``traditional'' equivalence theorems is that now ours' does not involve the Fisher information matrix.

\begin{theorem} \label{thm1}  Suppose that the conditions of Theorem \ref{prop1} hold and
 the infimum in \eqref{f1F2} and \eqref{f2F1} is attained at a unique point $\theta_2^* \in \Theta_2$ for the optimal design $\xi^*$.
\\
(a)
A design  $\xi^*$ is a  semi-parametric optimal design  for discriminating between   the model $f_1(y,x, { \overline\theta_1})$ and the class $\mathcal{F}_2$
if and only if the following inequality holds for all $x \in {\cal X}$:
\begin{equation} \label{equiv1}
I_{ 1,2}(x,f_1,f_2^*,{ \overline\theta_1},\theta_2^*) -  \int_{\cal X} I_{ 1,2}(x,f_1,f_2^*,{ \overline\theta_1},\theta_2^*) \, \xi^*(dx)\leq~0,
\end{equation}
with equality at the support points of $\xi^*$.  Here $I_{ 1,2}(x,f_1,f_2,{ \overline\theta_1},\theta_2)$ is defined in \eqref{KL1},
\begin{align*}
\theta_2^* &= \arginf_{\theta_2 \in \Theta_2} \int_{\cal X} I_{ 1,2}(x,f_1,f_2^*,{ \overline\theta_1},\theta_2) \, \xi^*(dx), \\
f_2^*(y,x,\theta_2)& = \frac{f_1(y,x,{ \overline\theta_1})}{1+\lambda(y-\eta_2(x,\theta_2))}
\end{align*}
and  $\lambda$ is found from~(\ref{eq:lambda}). Moreover, there is equality in
\eqref{equiv1} for all support point of $\xi^*$. \\
(b)
A design  $\xi^*$ is a  semi-parametric optimal design  for discriminating between   the model $f_2(y,x, { \theta_2})$ and the class $\mathcal{F}_1$
if and only if the following inequality holds for all $x \in {\cal X}$:
\begin{equation} \label{equiv2}
I_{ 1,2}(x,f_1^*,f_2,{\overline\theta_1,\theta_2^*}) -   \int_{\cal X}I_{ 1,2}(x,f_1^*,f_2,{ \overline\theta_1,\theta_2^*}) \, \xi^*(dx) \leq~0,
\end{equation}
with equality at the support points of $\xi^*$.  Here
\begin{align*}
{\theta_2^*} &= \arginf_{ \theta_2\in \Theta_2} \int_{\cal X} I_{ 1,2}(x,f_1^*,f_2,{\overline\theta_1,\theta_2^*}) \, \xi^*(dx), \\
f_1^*(y,x,\overline\theta_1) &= \frac{f_2(y,x,\theta_2) \exp(-\lambda y)}{\int f_2(y,x,\theta_2) \exp(-\lambda y) dy}
\end{align*}
and  $\lambda$ is found from~(\ref{eq:lambda2}). Moreover, there is equality in
\eqref{equiv2} for all support point of $\xi^*$.
\end{theorem}

\begin{example} \label{examotsu}  {\rm
This example illustrates the application of the equivalence theorem using an example from
\cite{Otsu2008}. The goal was to construct
 semi-parametric
optimal  designs for discriminating between $f_1(y,x, \overline\theta_1)$, the truncated normal density on the
  interval $[\eta_1(x,\overline\theta_1) - 3, \eta_1(x,\overline\theta_1) + 3]$ with mean $\eta_1(x,\overline\theta_1)$ and variance $1$,  and densities in the class $\mathcal{F}_2$.
  The two competing models defined on the same  design space is  ${\cal X} = {[-1,1]}$ are
\begin{eqnarray} \label{mod1}
\eta_1 (x,\theta_1) &=& \theta_{1,1} + \theta_{1,2} e^x + \theta_{1,3} e^{-x}, ~\\ 
\label{mod2}
\eta_2(x,\theta_2)  &=& \theta_{2,1} + \theta_{2,2} x + \theta_{2,3} x^2,
\end{eqnarray}
where  $\overline\theta_1 = (4.5, -1.5, -2)$. \cite{Otsu2008} determined the
 semi-parametric
optimal  design  for discriminating between   the model $f_1(y,x, \overline\theta_1)$ and the class $\mathcal{F}_2$
as
\begin{align} \label{otsudesign}
\tilde \xi =
\begin{bmatrix}
-1.000 & -0.266 & 0.721 & 1.000 \\
0.377 & 0.198 & 0.244 & 0.181
\end{bmatrix}.
\end{align}
 The left part of Figure \ref{eqivotsu} shows the plots of the sensitivity function on the left-hand side of \eqref{equiv1} for the design found by \cite{Otsu2008} (left). 
 We conclude  that this design is not
optimal    for discriminating between   the model  $f_1(y,x, \overline\theta_1)$ and the class $\mathcal{F}_2$, because the function is  positive throughout the design space. The  subfigure on the right shows the sensitivity function of the     design
\begin{align} \label{tdesign}
 \xi^* =
\begin{bmatrix}
-1.000 & -0.670 & 0.142 & 0.959 \\
0.253 & 0.428 & 0.247 & 0.072
\end{bmatrix},
\end{align}
 which has been found maximizing the criterion provided by Theorem \ref{prop1} and   confirms its optimality.
\begin{figure}[H]
\label{eqivotsu}
\centering
\subfigure{%
\includegraphics[width=70mm]{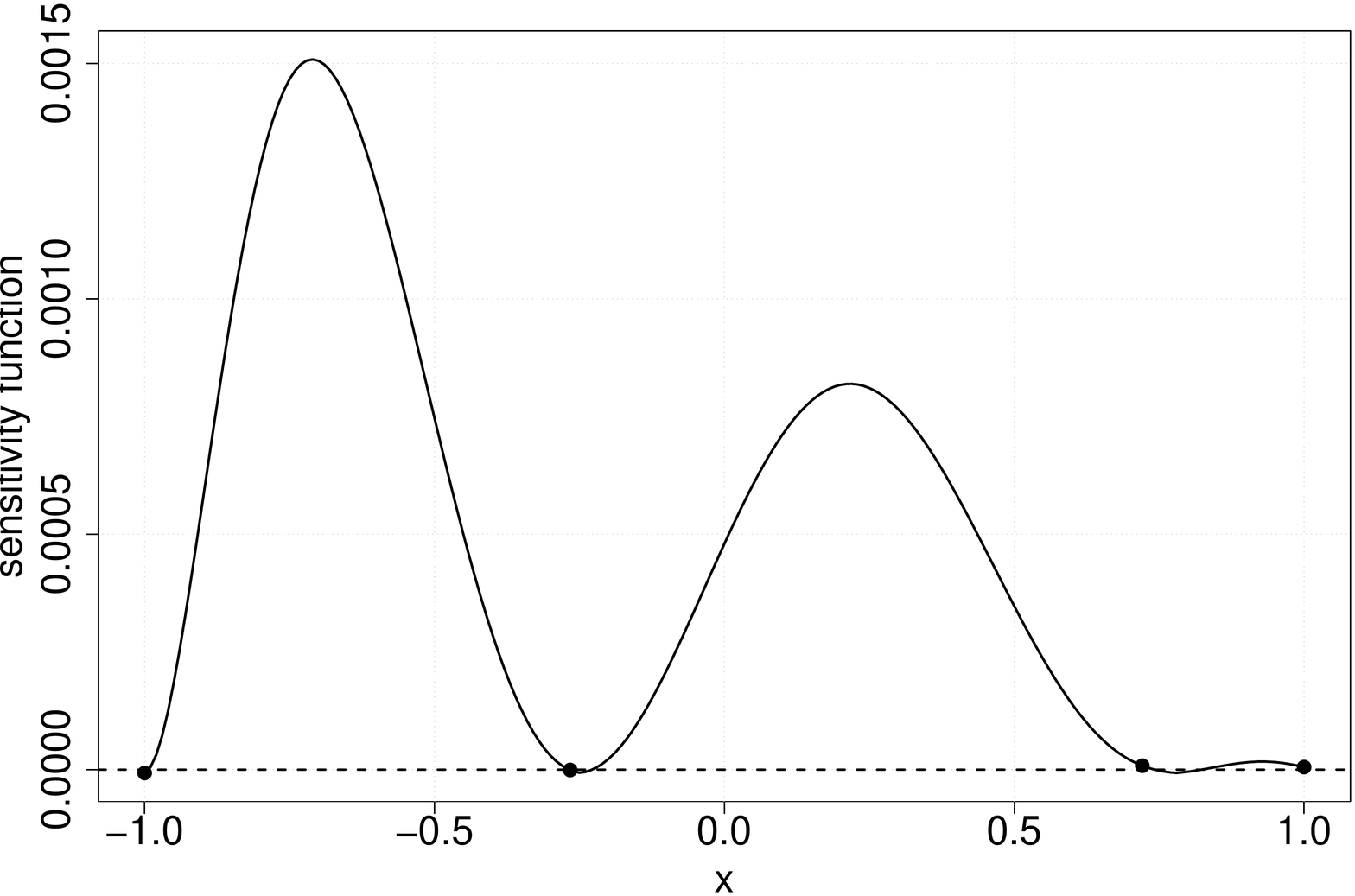}
}
\subfigure{%
\includegraphics[width=70mm]{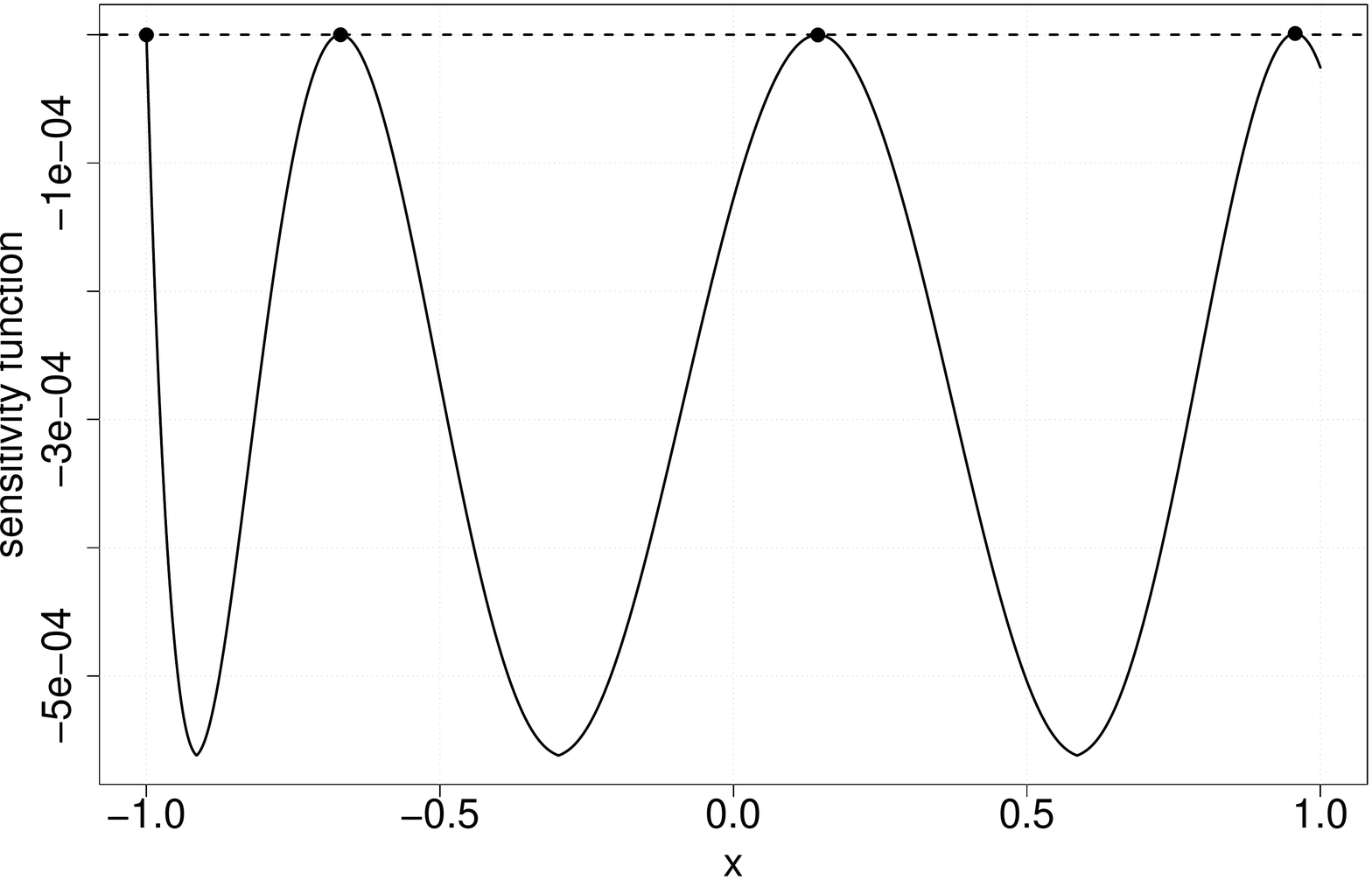}
}
\caption{\it The sensitivity functions of two designs for discriminating between  models
\eqref{mod1} and \eqref{mod2}.  The left figure is for the design \eqref{otsudesign} calculated by \cite{Otsu2008} and the right figure is for the $T$-optimal  design  \eqref{tdesign}.}
\end{figure}
The values of the criterion \eqref{f1F2} for different designs are given by   $K_{(a)}  (\tilde \xi) = 3.27779 * 10^{-6}$ and  $K_{(a)}  ( \xi^*) = 0.0005580455$
 demonstrating that $\xi^*$
performs substantially better than $\tilde \xi$; in fact, the efficiency of $\tilde \xi$ of the design determined by \cite{Otsu2008}
relative to  $ \xi^*$
is  approximately $ 0.006$. It turns out that the design in (\ref{tdesign}) is also  the $T$-optimal design.
In the next section, we prove in  Theorem \ref{thm4}  that for models \eqref{mod1} and \eqref{mod2}, the
semi-parametric
optimal  design  for discriminating between   the model $f_1(y,x, \overline\theta_1)$ and the class $\mathcal{F}_2$ actually coincides
with the  $T$-optimal design proposed by \cite{atkfed1975a}.
}
\end{example}


\section{Analytical results} \label{sec4}
\def\theequation{4.\arabic{equation}}
\setcounter{equation}{0}

In this section we investigate  relationships between  the semi-parametric optimality criteria considered
in this paper  and the optimality criteria for model discrimination with normal errors [see \cite{atkfed1975a}  and  \cite{ucibog2005}].
A  "classical"  $T$-optimal design,  say $\xi_T^*$,
for discriminating between two models
maximizes the criterion
\begin{equation} \label{topt}
 \inf_{\theta_2 \in \Theta_2} \int_{\cal X} [\eta_1(x,\overline\theta_1) - \eta_2(x,\theta_2)]^2 \xi(dx)
\end{equation}
among all designs on the design space [see \cite{atkfed1975a}], where $\overline\theta_1$ is fixed. We will presume throughout this section that the infimum in~\eqref{topt} is attained at a unique point $\theta_2^*$ when $\xi = \xi^*_T$. It follows by similar arguments as given in  \cite{wiens2009} that under the assumption of a normal
 distribution, the power of the likelihood ratio test for the hypotheses
 \begin{equation} \label{hypmeant}
 H_0:  \eta(x) = \eta_1(x,\overline\theta_1) ~~\mbox{versus} ~~ H_1:   \eta(x) =\eta_2(x,\theta_2)  ~~\mbox{for some } \theta_2 \in \Theta_2
\end{equation}
is an increasing function of the criterion \eqref{topt}. The following results give a
sufficient condition for the $T$-optimal discriminating design to be
a semi-parametric optimal  design  in the sense of Section \ref{sec2}.

\begin{theorem} \label{thm4}
Let the assumptions of the Theorem~\ref{prop1} (a) be fulfilled and assume further that the density  $f_1(y,x,\overline\theta_1)$ can be represented in the form
\begin{equation} \label{symass}
f_1(y,x,\overline\theta_1)  = g(y - \eta_1(x,\overline\theta_1))
\end{equation}
where $g$ is a symmetric  density function   supported in the interval  $[-a,a]$, i.e.  $f_1$ has support $[-a+\eta_1(x,\overline\theta_1), a + \eta_1(x,\overline\theta_1)]$.  The $T$-optimal discriminating design maximizing the criterion \eqref{topt} is  a  semi-parametric optimal  design for discriminating between  the model $f_1(y,x, \overline\theta_1)$ and the class $\mathcal{F}_2$.
 \end{theorem}

\medskip

A similar result is available for the   semi-parametric
optimal  designs for discriminating between  the model $f_2(y,x, \theta_2)$ and the class $\mathcal{F}_1$. For this
purpose we consider the situation, where
 $f_2(y,x,\theta_2) $ is the  density of the normal distributions ${\cal N}  (\eta_2(x,\theta_2) , v^2_2(x,\theta_2))$.
 If $f_1(y,x,\overline\theta_1) $ is also the density of a  ${\cal N}  (\eta_1(x,\overline\theta_1) , v^2_2(x,\theta_2))$
  it can be shown that
the  power of the likelihood ratio test for the hypotheses    \eqref{hypmeant}
  is an increasing function of the quantity 
\begin{align} \label{topthet}
\inf_{\theta_2 \in \Theta_2} \int_{\cal X} \frac{[\eta_1(x,\overline\theta_1) - \eta_2(x,\theta_2)]^2}{v^2_2(x,\theta_2)} \xi(dx).
\end{align}
The maximization of this expression corresponds to the KL-optimal design problem for discriminating between two normal
distributions with the same variance as considered by \cite{loptomtra2007}. Our following result shows that this design
 is   also optimal for discriminating between  the model $f_2(y,x, \theta_2)$ and the class $\mathcal{F}_1$.

\begin{theorem} \label{thm5}
Assume that  $f_2(y,x,\theta_2)$ is the  density  of a normal distribution  with mean $\eta_2(x,\theta_2)$ and variance $v_2^2(x,\theta_2)$.  The optimal design maximizing \eqref{topthet} is  a  semi-parametric
optimal  design for discriminating between  the model $f_2(y,x, \theta_2)$ and the class $\mathcal{F}_1$ and vice versa.
Moreover, the best approximation
$f_1^*(y,x,\overline\theta_1)$ is  a normal density with mean $\eta_1(x,\overline\theta_1)$ and variance $v_2^2(x,\theta_2)$.
\end{theorem}

\section{Numerical results}
\label{sec5}
\def\theequation{5.\arabic{equation}}
\setcounter{equation}{0}

The numerical construction of semi-parameteric optimal discrimination designs  is a very challenging problem. In 
 this section we describe techniques for finding semi-parametric optimal designs and illustrate our approach in two examples.
It follows from the results of Section \ref{sec2}   that the first step in the determination of the optimal designs consists in an efficient  solution  of the 
 equations  \eqref{eq:lambda}   and \eqref{eq:lambda2}. In a second step  any numerical method  for the determination of    KL-optimal discrimination
 designs can be adapted to the minimax problems obtained from   Theorem \ref{prop1} as the
representations  \eqref{f1F2symp} and  \eqref{f2F1symp} have a similar  structure as the KL-optimality criteria considered in \cite{loptomtra2007}. 
Even the second step defines a  very challenging problem and some recent results and algorithms for KL-optimality criteria
 can be found in \citep{stegmaier}, \cite{BraessDette2013,detmelguc2015} and \cite{detmelguc2016}. As the focus in this paper is on the new semi-parametric design criteria we concentrate on the 
 first step in the following discussion. For the second step we  used  an adaptation  of  the first-order algorithm of  \cite{atkfed1975a}, because it can easily be implemented. 
 
Let  $\delta$ be a user-selected positive constant. For finding the numerical solution of the equation
\eqref{eq:lambda}  we propose the following algorithm: 
 \begin{itemize}
 \item if  $\eta_1 (x,\overline\theta_1)  = \eta_2  (x,\theta_2)$,  set $\lambda = 0$;
 \item   if $\eta_1  (x,\overline\theta_1) < \eta_2 (x,\theta_2)$,  choose a solution in the interval ${{\Lambda}^-=[-1/(y_{x,\max}-\eta_2(x,\theta_2)), -\delta]}$;
 \item   if $\eta_1  (x,\overline\theta_1) > \eta_2 (x,\theta_2)$,  choose
 a solution in  the interval ${{\Lambda}^+=[\delta, -1/(y_{x,\min}-\eta_2(x,\theta_2))]}$.
 \end{itemize}
Similarly, the solution of \eqref{eq:lambda2} can be obtained as follows.
We search for $\lambda > 0$ if $\eta_1(x,\overline\theta_1) < \eta_2(x,\theta_2)$ so that $\lambda$  shifts the predefined density $f_2(y,x,\theta_2)$ to the left and,  search for $\lambda < 0$ if $\eta_1(x,\overline\theta_1) > \eta_2(x,\theta_2)$.
Let  $\beta$ be a  large user-selected positive constant, let $\delta$ be a small positive constant and  assume that the solution of \eqref{eq:lambda2} is in $[-\beta,+\beta]$. We note that
\begin{align*}
\mu^{\prime}(\lambda) f_2(y,x,\theta_2) \exp(-\lambda y) > 0, \; \forall \lambda \in [-\beta, +\beta],\end{align*}
 for all $y \in
 {\cal S}_{f_2, \theta_2, x},$
where $\mu^\prime(\lambda)$ is defined in~\eqref{f2mu}. We suggest the following algorithm  for finding a numeral solution of~\eqref{eq:lambda2}:
\begin{itemize}
\item if  $\eta_1 (x,\overline\theta_1)  = \eta_2  (x,\theta_2)$, set $\lambda = 0$;
 \item   if $\eta_1  (x,\overline\theta_1) < \eta_2 (x,\theta_2)$,  choose   a solution in the interval ${\Lambda}^+ = [+\delta, +\beta]$;
 \item   if $\eta_1  (x,\overline\theta_1) > \eta_2 (x,\theta_2)$, choose
 a solution in  the interval $\Lambda^- = [-\beta,-\delta]$.
\end{itemize}

\begin{figure}
\centering
\subfigure[$T$-optimal]{%
\includegraphics[width=50mm]{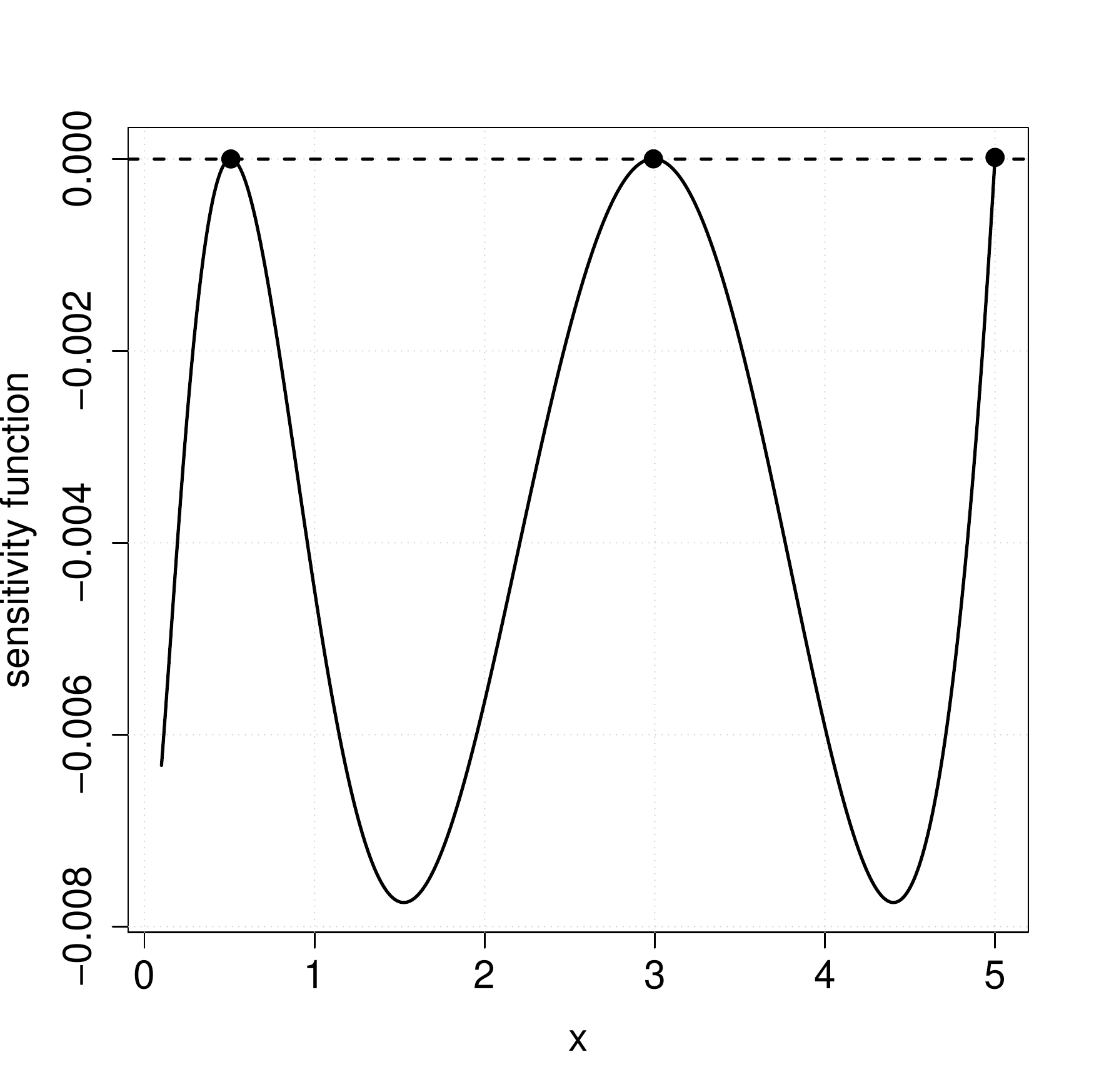}
}
\subfigure[$KL$-optimal]{%
\includegraphics[width=50mm]{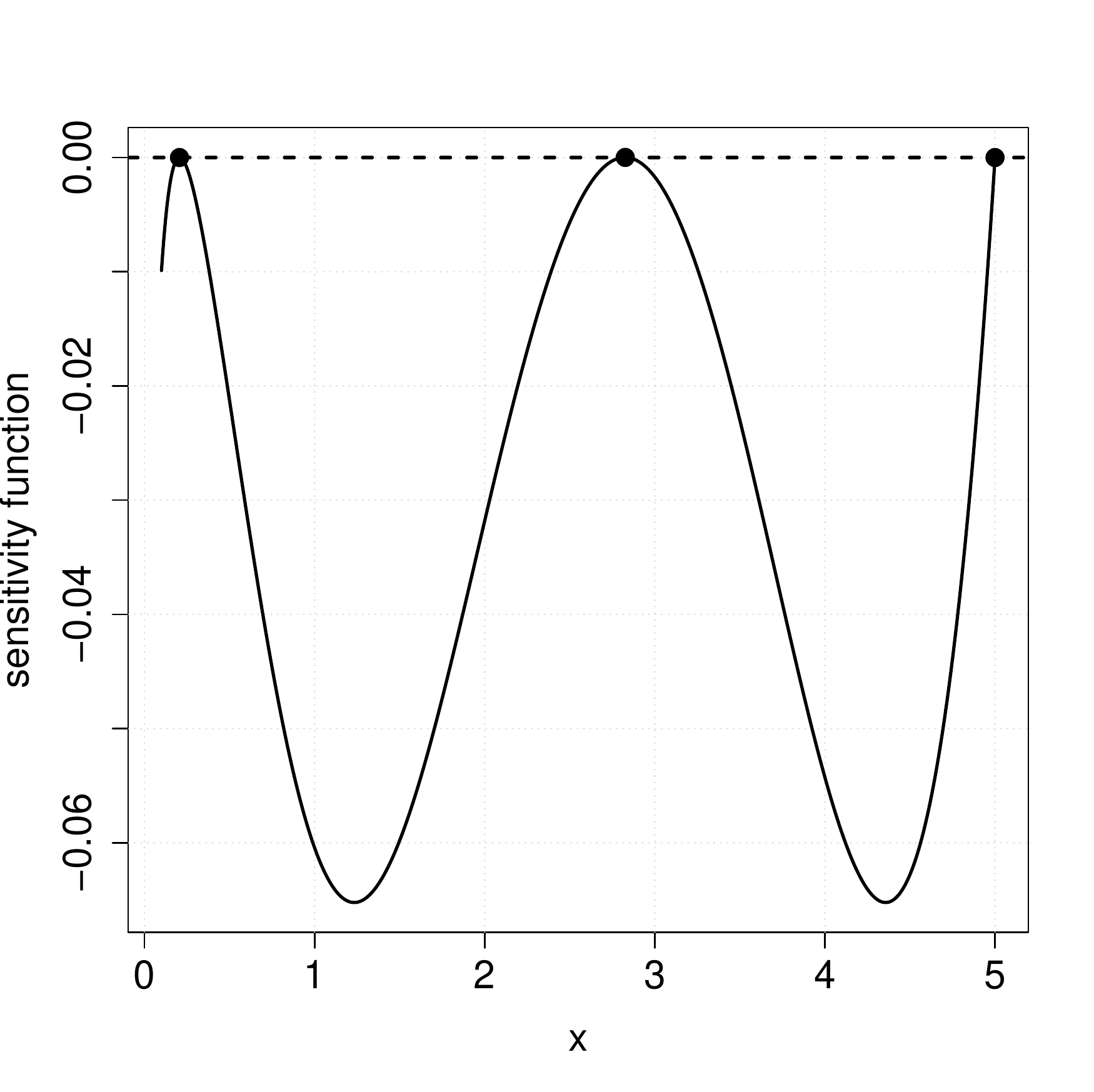}
}
\subfigure[$SKL$-optimal]{%
\includegraphics[width=50mm]{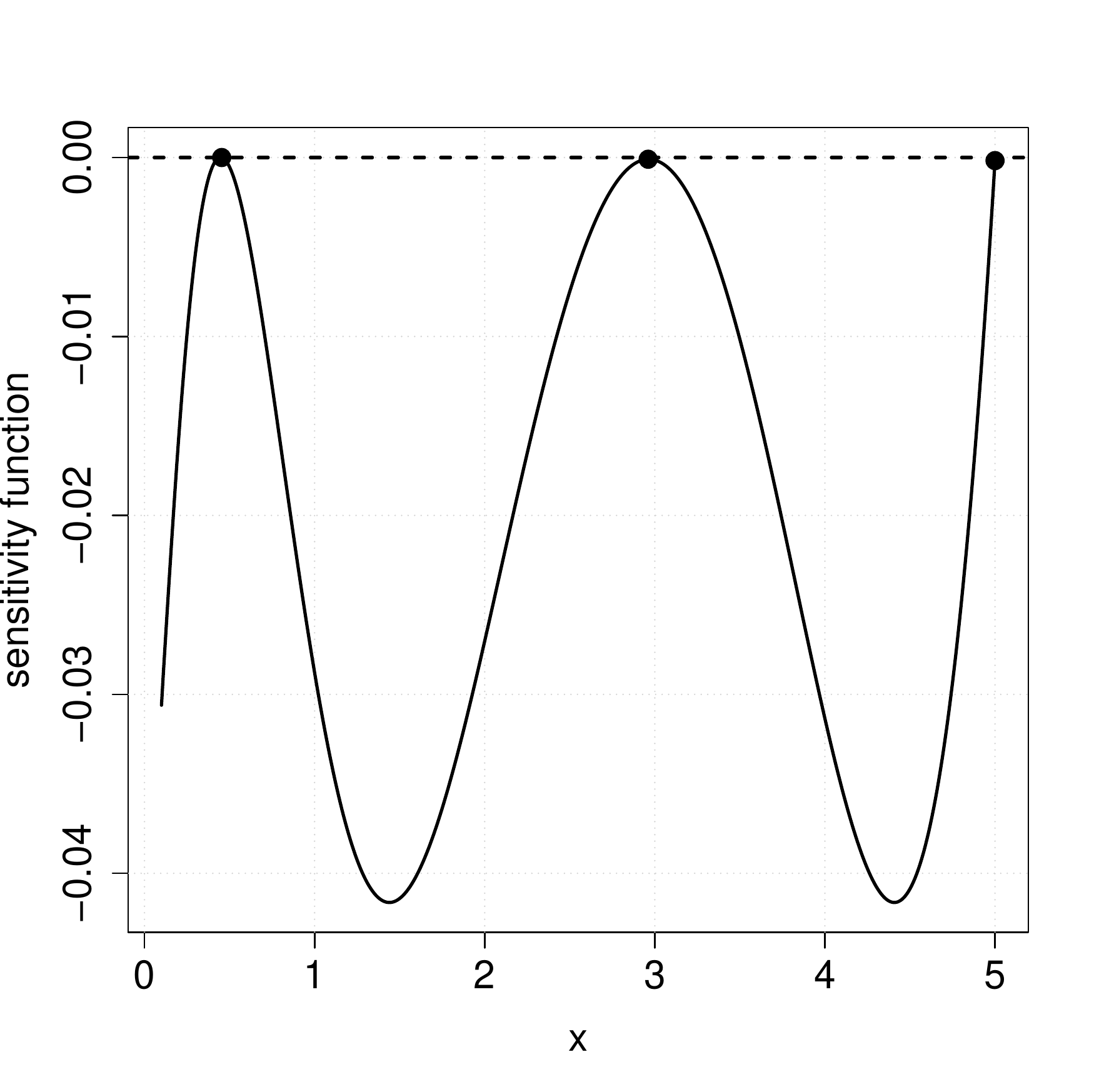}
}
\caption{\it  The sensitivity functions of the $T$-, $KL$- and semi-parametric $KL$-optimal discrimination (SKL) designs   in Table \ref{tab11}.}
\label{fig1}
\end{figure}

We now present two examples, where the   $T$-optimal and semi-parametric $KL-$optimal designs are  determined numerically and  shown to be
different.
To be precise, consider the optimal design problem from ~\cite{loptomtra2007}, where they were interested to discriminate between the two models:
\begin{align} \label{mod11}
\eta_1(x,\theta_1) &= \theta_{1,1} x + \frac{\theta_{1,2} x}{x + \theta_{1,3}}, \\ 
\eta_2(x,\theta_2) &= \frac{\theta_{2,1} x}{x + \theta_{2,2}}. \label{mod21}
\end{align}
The design space for both models is the interval $[0.1,5]$ and we assume that the first model has fixed parameters $\overline{\theta}_1 = (1,1,1)$.  We construct three different types of optimal discrimination designs for this problem: a $T$-optimal design, a KL-optimal design for lognormal errors (with fixed variances $v^2_1(x,\overline\theta_1) = v^2_2(x,\theta_2) = 0.1$) and a semi-parametric KL-optimal design (case a)) for a mildly truncated lognormal  density $f_1(y,x,\overline\theta_1)$ with location and scale parameters, respectively, given by
\begin{align*}
\mu_1(x,\overline\theta_1) &= \log\left[ \eta_1(x,\overline\theta_1) \right] - \frac{1}{2}\log\left[ 1+v^2_1(x,\overline\theta_1)/\eta_1^2(x,\overline\theta_1) \right], \\
\sigma^2_1(x,\overline\theta_1) &= \log\left[ 1+v^2_1(x,\overline\theta_1)/\eta_1^2(x,\overline\theta_1) \right].
\end{align*}
 The range for this density is the interval from $Q(0.0001,x,\overline\theta_1)$ to $Q(0.9999,x,\overline\theta_1)$, where $Q(p,x,\overline\theta_1)$ is quantile function of ordinary lognormal density with mean $\eta_1(x,\overline\theta_1)$ and variance $v_1^2(x,\overline\theta_1) = 0.1$. We note that because of mild truncation $\eta_1(x,\overline\theta_1)$ does not correspond exactly to the mean of $f_1(y,x,\overline\theta_1)$ but is very close to it. The optimal
 discrimination designs under the three different criteria are displayed  in Table \ref{tab11}, where we also
 show  the ``optimal''  parameter $\theta_2^*$ of  the second model corresponding to the minimal value with respect to the parameter $\theta_2$.
 We observe substantial differences between the $T$, $KL$- and semi-parametric $KL$-optimal discrimination (SKL) designs.
Figure \ref{fig1} displays the sensitivity  functions of the three designs  and  confirms the   optimality of each of the designs.

\begin{figure}
\centering
\includegraphics[width=55mm]{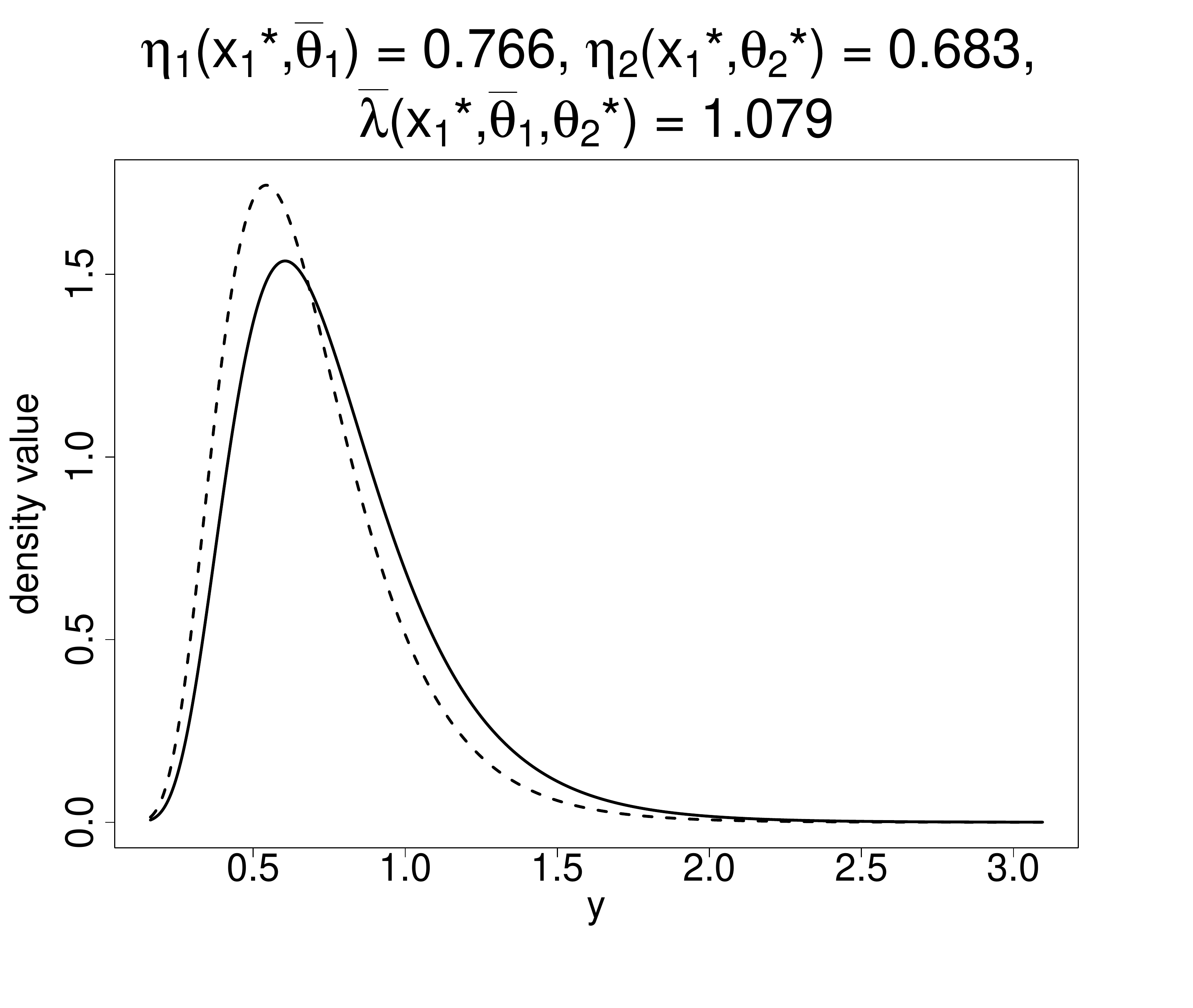}
\includegraphics[width=55mm]{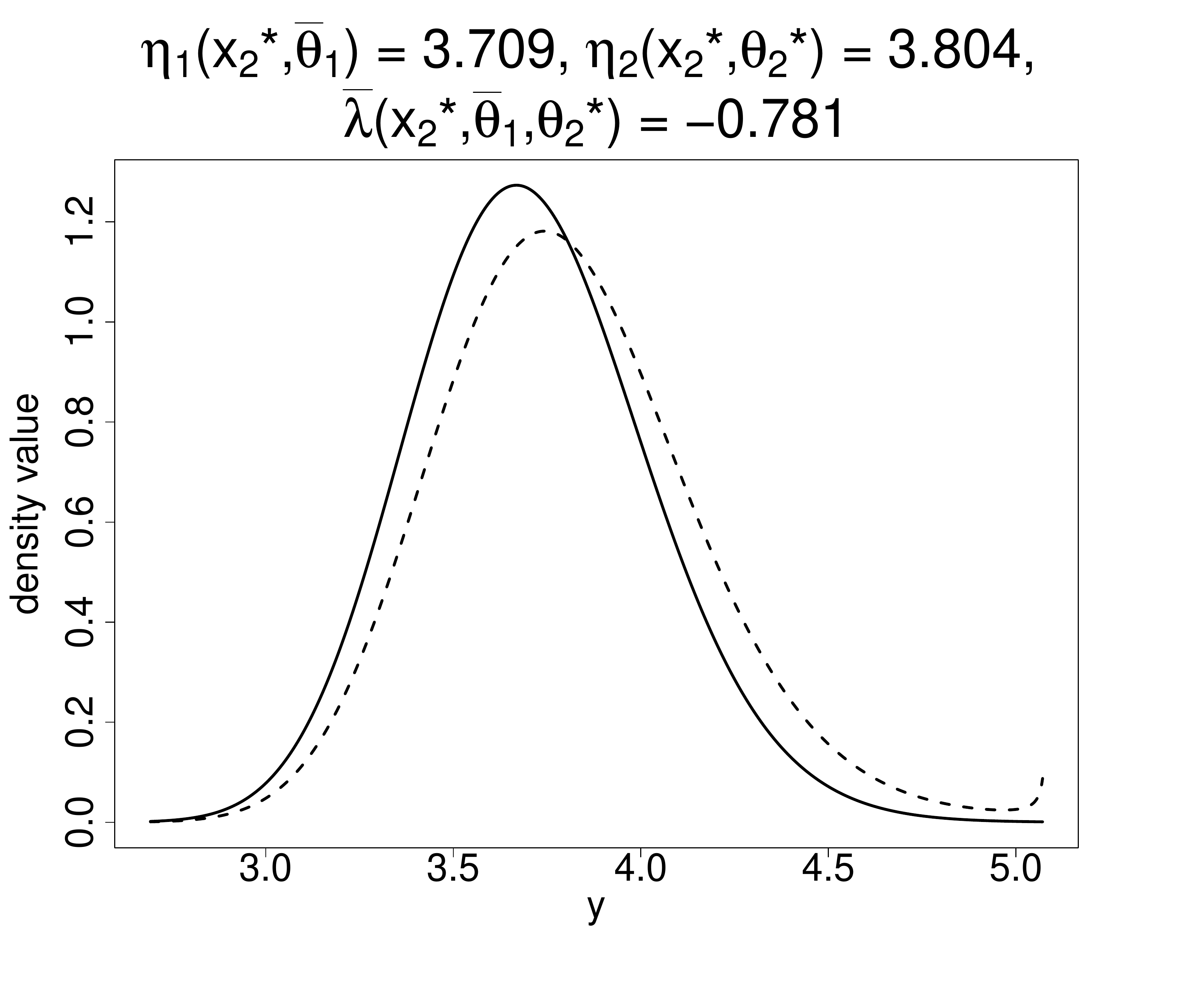}
\includegraphics[width=55mm]{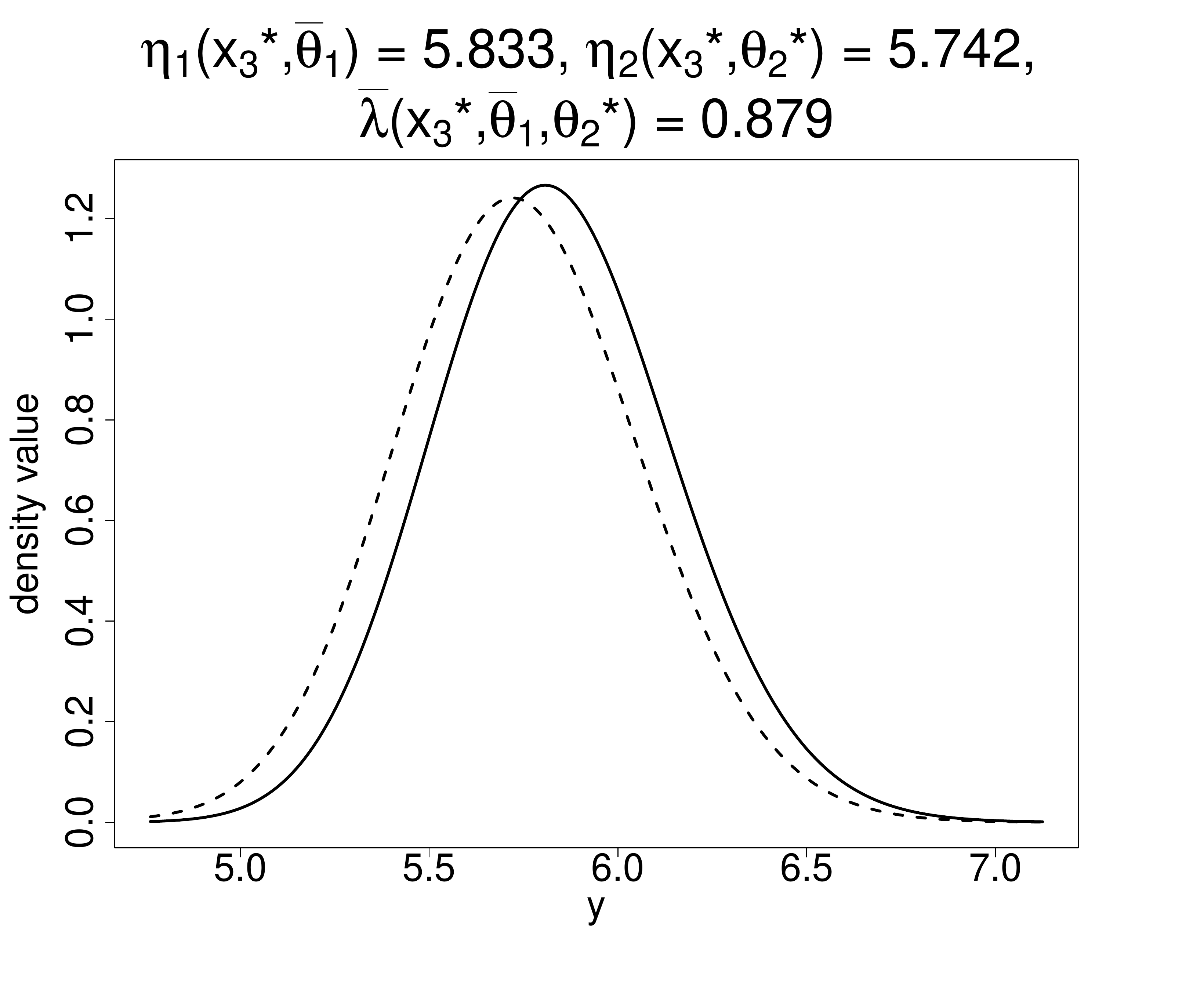}
\caption{\it The density $f_1(y,x_i^*,\overline\theta_1)$ of a truncated lognormal distribution (solid line) on the interval $[Q(0.0001,x_i^*,\overline\theta_1),Q(0.9999,x_i^*,\overline\theta_1)]$ and the corresponding ``optimal'' density $f_2^*(y,x_i^*,\theta_2^*)$, defined by~\eqref{f1F2opt}, for the support points $x_i^*$, $i = 1,2,3$ of SKL-optimal design and parameter values $\theta_2^*$ from Table~\ref{tab11}.}
\label{fig2:figure}
\end{figure}

Table \ref{tab21}  displays the efficiencies of $T$, $KL$- and semi-parametric $KL$-optimal discrimination (SKL) designs with respect to the different
 criteria. For example, the value $0.247$ in the first row  is  the efficiency of the $KL$-optimal design with respect
 to the $T$-optimality criterion. We observe that the $T$-   and  $KL$- optimal discrimination  design are not very
  robust under a variation of the criteria, where the $KL$-optimal discrimination design has slight advantages.
  On the other hand,  the
  semi-parametric $KL$-optimal discrimination  design  yields moderate  efficiencies (about  $75 \%$) with respect to  the $T$- and $KL$-optimality criterion.
  Figure \ref{fig2:figure} shows  the plots of the functions $f_1(y,x_i^*,\overline\theta_1)$ and $f_2^*(y,x^*_i,\theta_2^*)$ for the support points $x_i^*$, $i = 1,2,3$ of $SKL$-optimal design. Above each figure the values $\eta_1(x_i,\overline\theta_1)$, $\eta_2(x_i,\theta_2^*)$ and  $\overline\lambda(x_i^*,\overline\theta_1,\theta_2^*)$, the solution of~\eqref{eq:lambda}, are presented.  We note that the densities $f_1$ and $f_2^*$ depend on the parameters $\overline \theta_1$ and $\theta_2^*$ only through $\eta_1(x, \overline \theta_1)$ and $\eta_2(x, \theta^*_2)$.

\begin{table}
\centering
\begin{tabular}{ |l|c|c| }
\hline
Design type & $\xi^*$ & $\theta_2^*$ \\
\hline
$T$-optimal &
$
\begin{matrix}
0.508 & 2.992 & 5.000 \\
0.580 & 0.298 & 0.122
\end{matrix}
$ &
$(22.564, 14.637)$ \\
\hline
$KL$-optimal &
$
\begin{matrix}
0.206 & 2.826 & 5.000 \\
0.574 & 0.308 & 0.118
\end{matrix}
$ &
$(20.552, 12.962)$ \\
\hline
$SKL$-optimal &
$
\begin{matrix}
0.454 & 2.961 & 5.000 \\
0.531 & 0.344 & 0.125
\end{matrix}
$ &
$(22.045, 14.197)$ \\
\hline
\end{tabular}
\caption{\label{tab11}
\it  $T$, $KL$- and semi-parametric $KL$-optimal discrimination (SKL) designs for discriminating between models \eqref{mod11} and  \eqref{mod21}.}
\end{table}

\begin{table}
\centering
\begin{tabular}{|l|l|l|l|}
\hline
$K(\xi) \setminus \xi$ & $T$ & $KL$ & $SKL$ \\
\hline
$T$ & 1 & 0.247 & 0.741 \\
\hline
$KL$ & 0.653 & 1 & 0.787 \\
\hline
$SKL$ & 0.55 & 0.397 & 1 \\
\hline
\end{tabular}
\caption{\label{tab21}
\it   Efficiencies of the $T$, $KL$- and semi-parametric $KL$-optimal discrimination (SKL) designs for the models \eqref{mod11} and  \eqref{mod21}
under different optimality  criteria.}
\end{table}

\begin{figure}
\centering
\subfigure[$T$-optimal]{%
\includegraphics[width=50mm]{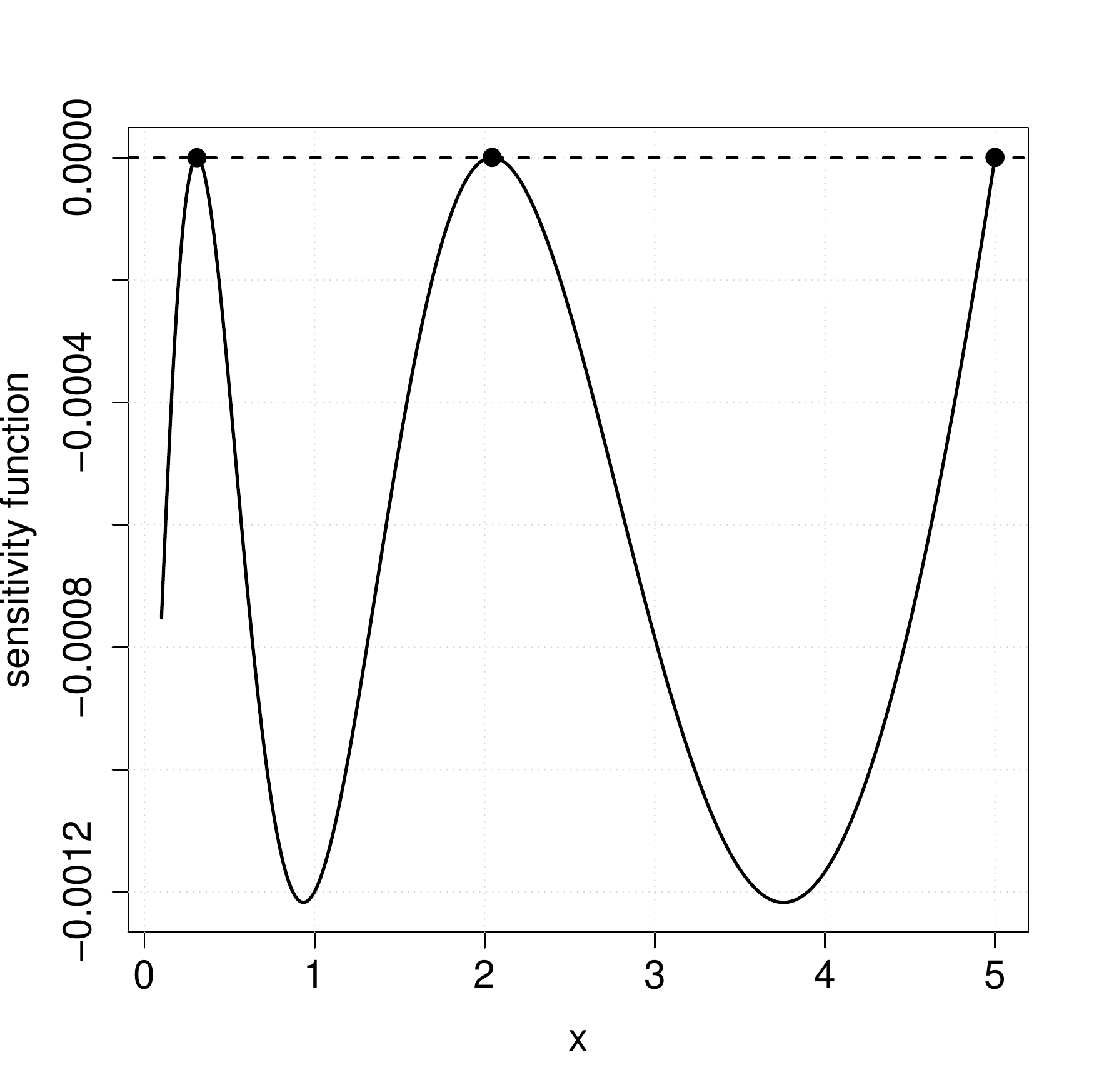}
}
\subfigure[$KL$-optimal]{%
\includegraphics[width=50mm]{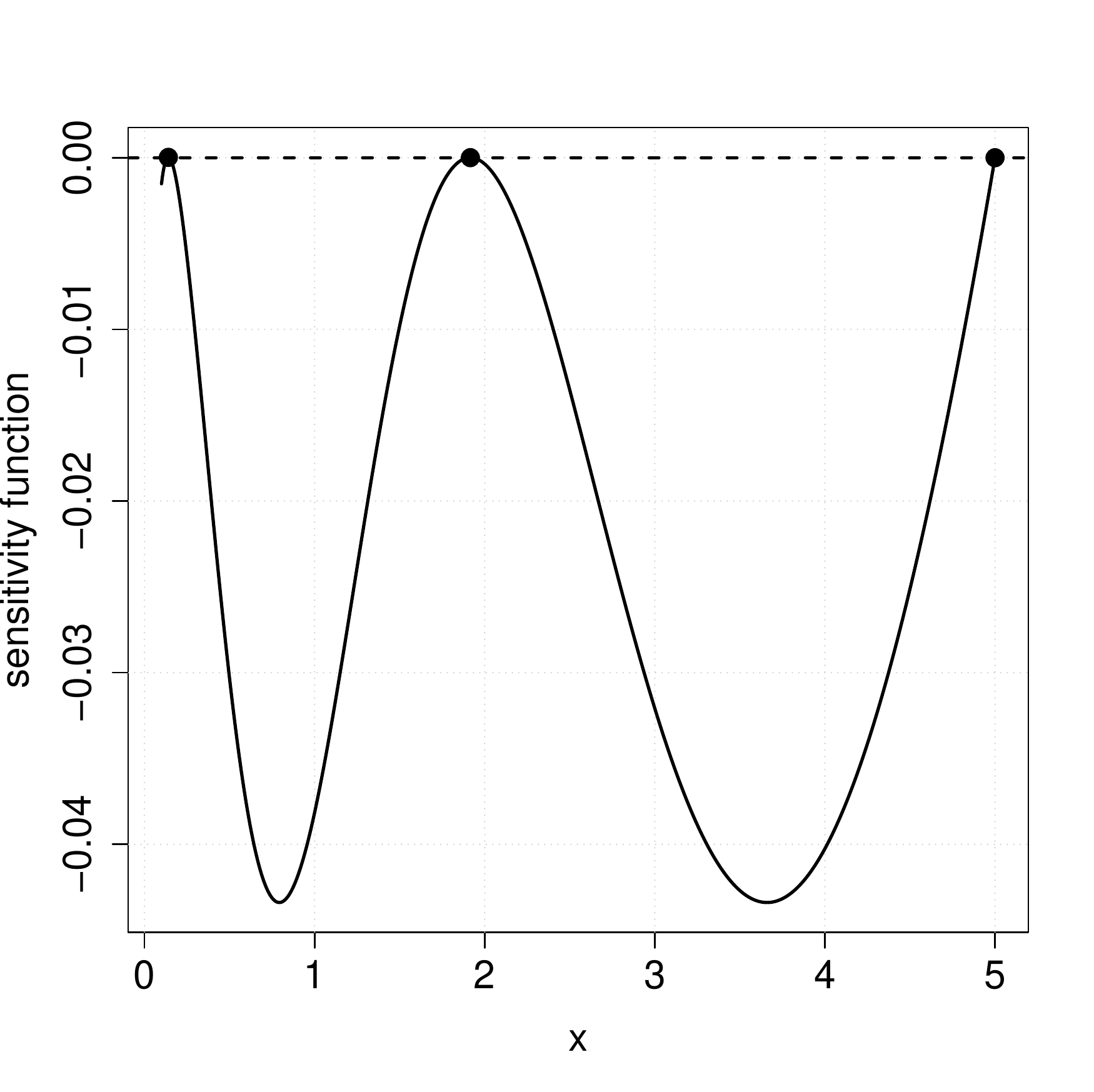}
}
\subfigure[$SKL$-optimal]{%
\includegraphics[width=50mm]{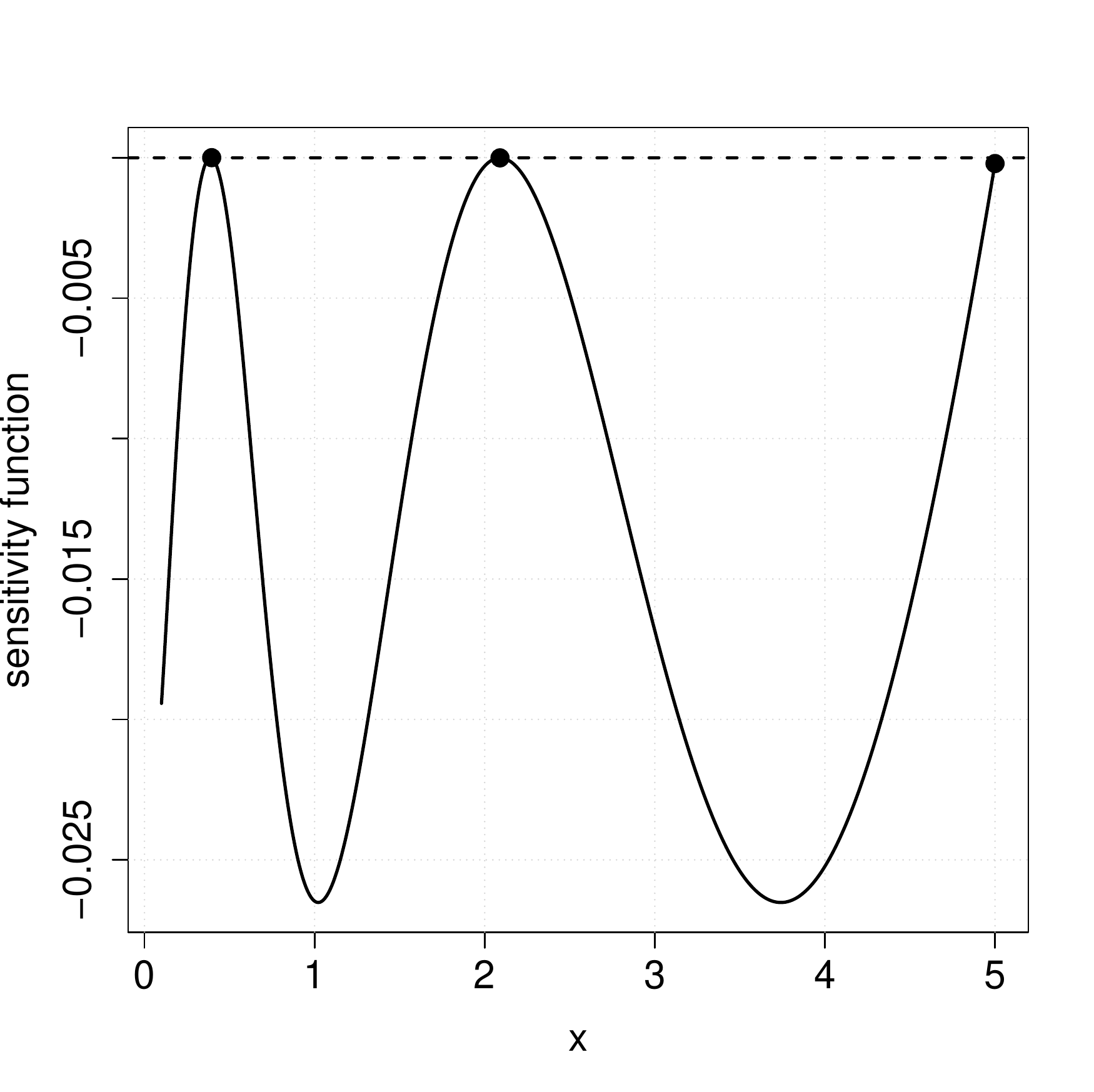}
}
\caption{\it  The sensitivity functions of the $T$-, $KL$- and semi-parametric $KL$-optimal discrimination (SKL) designs in Table \ref{tab3}.
 \label{fig32} }
\end{figure}

\begin{figure}
\centering
\includegraphics[width=55mm]{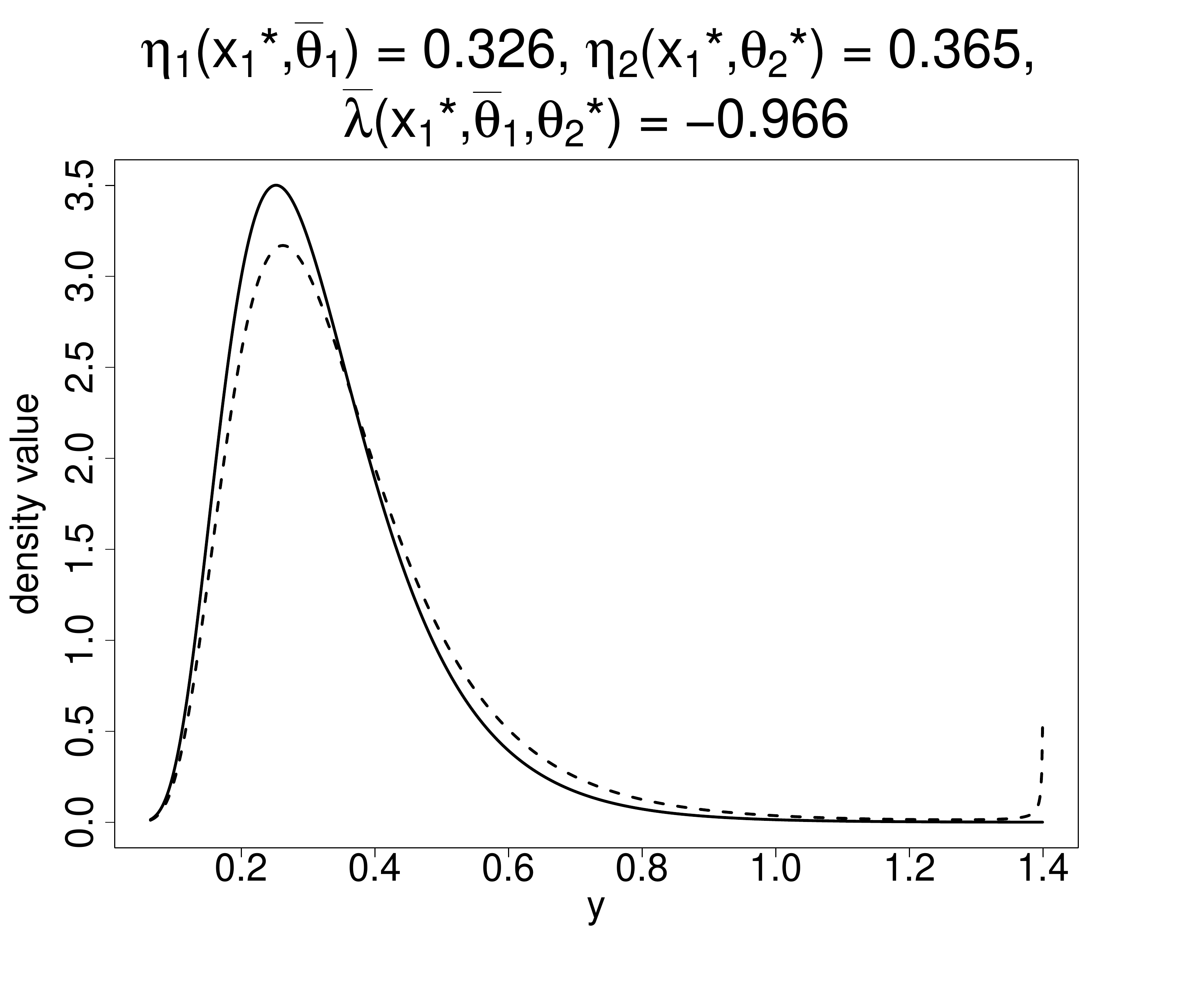}
\includegraphics[width=55mm]{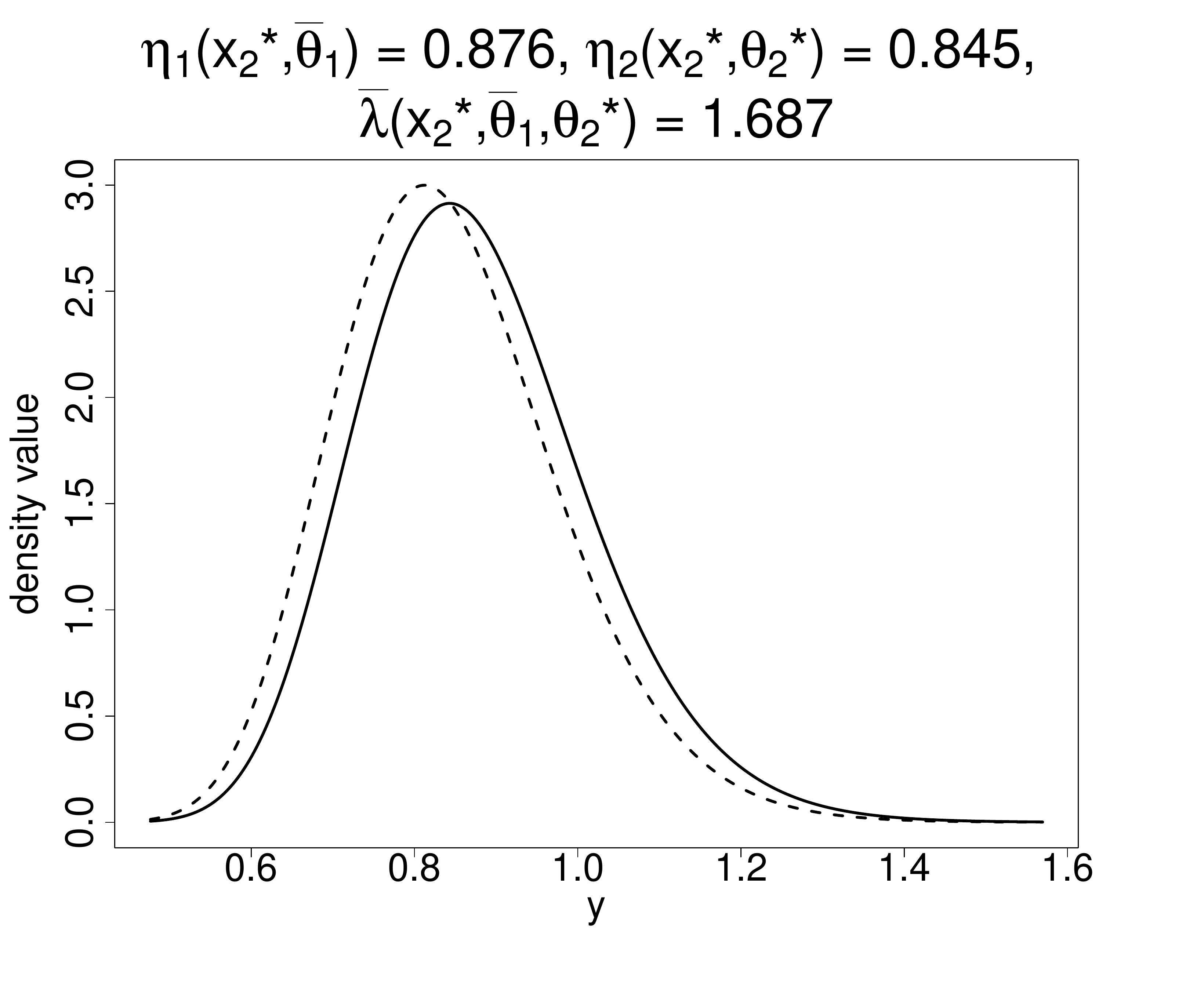}
\includegraphics[width=55mm]{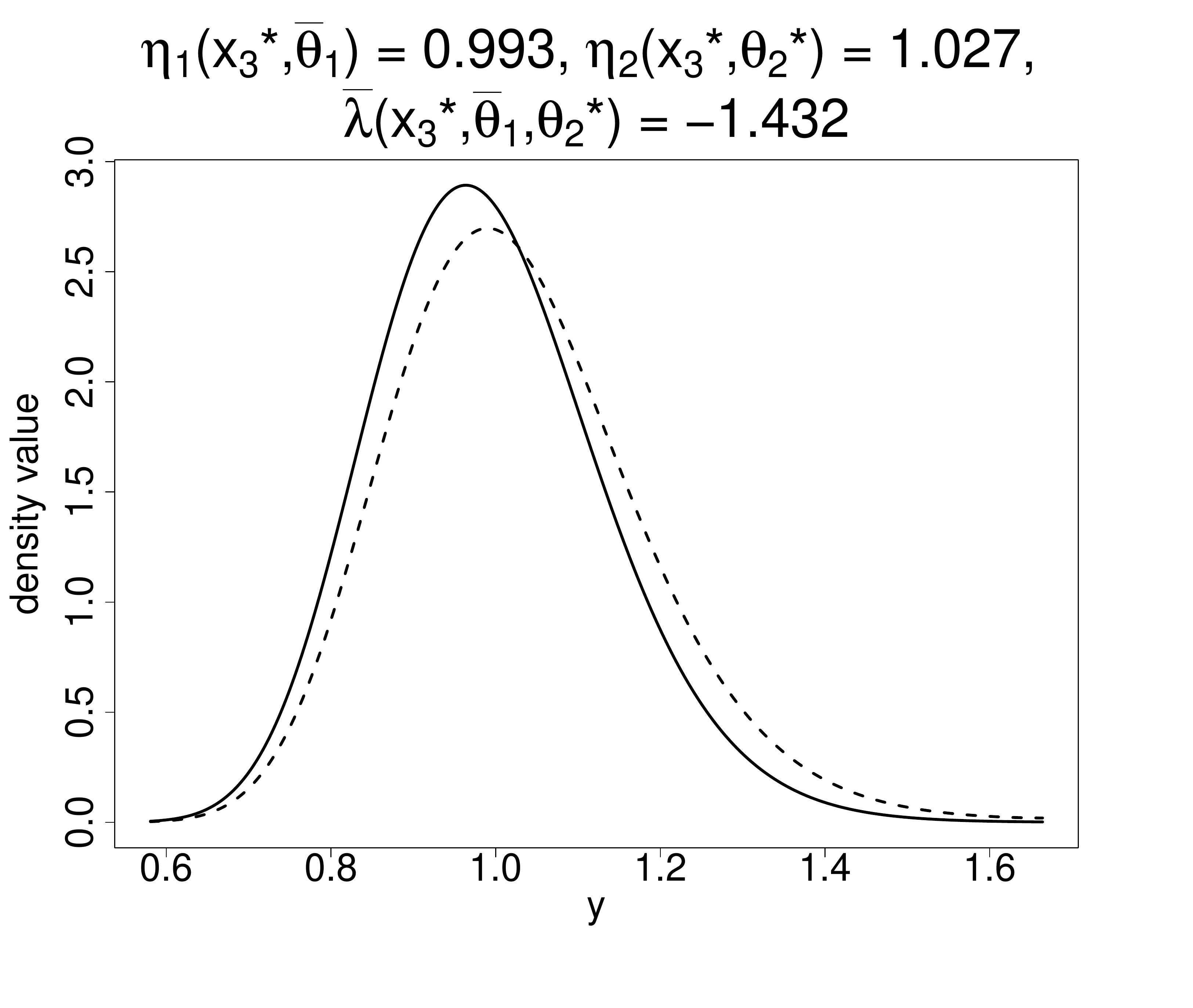}
\caption{\it The density $f_1(y,x_i^*,\overline\theta_1)$ of a truncated lognormal distribution (solid line) and the ``optimal'' density $f_2^*(y,x_i^*,\theta_2^*)$ for the support points $x_i^*$, $i = 1,2,3$ of SKL-optimal design and corresponding ``optimal'' parameter values $\theta_2^*$ from Table~\ref{tab3}.}
\label{fig3:figure}
\end{figure}

As a second example we now consider a similar problem with a different function $\eta_1(x,\theta_1)$ \citep{wiens2009}.  The two models of interest are
\begin{eqnarray}
 \label{mod1c}
\eta_1(x,\theta_1) &=& \theta_{1,1} \big \{ 1 - \exp(-\theta_{1,2} x) \big\}, \\ 
 \eta_2(x,\theta_2) &=&  \frac{\theta_{2,1} x}{\theta_{2,2}+ x}, \label{mod2c}
\end{eqnarray}
where the design space is given by  the interval  ${\cal X}  = [0.1,5]$. Here we fix the  parameters of the  model \eqref{mod1c}
as $\overline \theta_1 = (1,1)$ and determine the $T$-optimal, $KL$-optimal (for lognormal errors) and a semi-parametric $KL$-optimal
design (case a)) for mildly truncated lognormal errors. The error variances for the $KL$-optimal  discrimination
design are $v_1^2(x,\overline\theta_1) = v_2^2(x,\theta_2) =0.02$ and for the  semi-parametric $KL$-optimal  the variance is $v_1^2(x,\overline\theta_1) = 0.02$.
The optimal designs with respect to the different criteria are presented in Table  \ref{tab3},  where we also
 show  the corresponding  parameter $\theta_2^*$ of  the second model corresponding to the minimal value with respect to the parameter $\theta_2$.
 We observe again substantial differences between the optimal discrimination designs with respect to the different criteria, and a
 comparison of the efficiencies of the optimal designs with respect ot the different criteria  in Table \ref{tab32} shows a similar picture as in the first example.
 Figure \ref{fig32} displays the sensitivity functions of the $T$-, $KL$- and semi-parametric $KL$-optimal discrimination designs and the three subplots confirm optimality of the three three-point designs.
  Figure \ref{fig3:figure} shows  the plots of the functions $f_1(y,x_i^*,\overline\theta_1)$ and $f_2^*(y,x^*_i,\theta_2^*)$ for the support points $x_i^*$, $i = 1,2,3$ of $SKL$-optimal design from Table~\ref{tab3}, the legend above each plot has the same meaning as before.

\begin{table}
\centering
\begin{tabular}{ |l|c|c| }
\hline
Design type & $\xi^*$ & $\theta_2^*$ \\
\hline
$T$-optimal &
$
\begin{matrix}
0.308 & 2.044 & 5.000 \\
0.316 & 0.428 & 0.256
\end{matrix}
$ &
$(1.223, 0.948)$ \\
\hline
$KL$-optimal &
$
\begin{matrix}
0.140 & 1.916 & 5.000 \\
0.333 & 0.402 & 0.465
\end{matrix}
$ &
$(1.242, 1.006)$ \\
\hline
$SKL$-optimal &
$
\begin{matrix}
0.395 & 2.090 & 5.000 \\
0.396 & 0.355 & 0.249
\end{matrix}
$ &
$(1.216, 0.920)$ \\

\hline
\end{tabular}
\caption{\label{tab3}
\it  $T$, $KL$- and semi-parametric $KL$-optimal discrimination (SKL) designs for discriminating between the models \eqref{mod1c} and  \eqref{mod2c}.}
\end{table}

\begin{table}
\centering
\begin{tabular}{|l|l|l|l|}
\hline
$K(\xi) \setminus \xi$ & $T$ & $KL$ & $SKL$ \\
\hline
$T$ & 1 & 0.360 & 0.663 \\
\hline
$KL$ & 0.835 & 1 & 0.655 \\
\hline
$SKL$ & 0.407 & 0.361 & 1 \\
\hline
\end{tabular}
\caption{\label{tab32}
\it   Efficiencies of the $T$, $KL$- and semi-parametric $KL$-optimal discrimination (SKL) designs for the models \eqref{mod1c} and  \eqref{mod2c} under various  optimality criteria.}
\end{table}

\newpage 

\section{Proofs} \label{sec6}
\def\theequation{6.\arabic{equation}}
\setcounter{equation}{0}

{\bf Proof of Theorem \ref{prop1}:}
We only prove the first part of the proposition. The second statement follows by similar arguments which are omitted for
the sake of brevity. We first introduce the following temporary notation
\begin{align}\label{not}
f_1(y,x,\overline\theta_1) = f_1(y), \; f_2(y,x,\theta_2) = f_2(y), \; \eta_1(x, \overline\theta_1) = \eta_1, \; \eta_2(x, \theta_2) = \eta_2,
\end{align}

and note that from  (2.6), we wish to minimize

\begin{align*}
I_{1,2} (x, f_1,f_2 ,  \theta_1, \theta_2)= \int  \log{ \left\{\frac{f_1(y)}{f_2(y)} \right\}} f_1(y)  dy
\end{align*}
with constraints
\begin{align}
\label{constr}
\int  f_2(y) dy = 1, \ ~\text{and}~ \int y f_2(y)dy = \eta_2.
\end{align}
Consequently, with the notation
\begin{align*}
H = \log{ \left\{\frac{f_1(y)}{f_2(y)} \right\}} f_1(y) + \mu f_2(y) + \lambda y f_2(y)~,
\end{align*}
 we obtain  the Euler-Lagrange equation
\begin{align*}
\frac{\partial H}{\partial f_2} = - \frac{f_1(y)}{f_2(y)} + \mu + \lambda y = 0
\end{align*}
and  the relation
\begin{align} \label{eq1}
f_2(y) = \frac{f_1(y)}{\mu + \lambda y}.
\end{align}
We may assume without loss of generality that  $\lambda \neq 0$ ; otherwise, we have $\mu = 1$ and $f_2(y) = f_1(y)$.
From the condition \eqref{constr} and \eqref{eq1} we obtain
\begin{align*}
\eta_2 &=
\int y f_2(y) dy  =
\int \frac{y}{\mu + \lambda y} f_1(y) dy\\
&  = \frac{1}{\lambda} \int \frac{\mu + \lambda y}{\mu + \lambda y} f_1(y) dy - \frac{\mu}{\lambda} \int \frac{1}{\mu + \lambda y} f_1(y) dy
= \frac{1}{\lambda} - \frac{\mu}{\lambda} .
\end{align*}
Hence we have $\mu = 1 - \eta_2 \lambda$ and
\begin{align*}
f_2(y) = \frac{f_1(y)}{1 + \lambda (y - \eta_2)} ~.
\end{align*}
Substituting the new expression for $f_2(y)$ into the remaining condition gives
\begin{align}
\label{eq:lambda1}
\int \frac{f_1(y)}{1+\lambda(y-\eta_2)} dy = 1.
\end{align}
By the assumption in the theorem, equation~\eqref{eq:lambda1} has a unique solution in the interval~\eqref{eq:lambda_bounds} and the inequality
\begin{align*}
1 + \overline\lambda (y - \eta_2) > 0
\end{align*}
holds for all $y \in \mathcal{S}_{f_1,\overline{\theta}_1 ,  x}$, where
$\overline\lambda$ solves \eqref{eq:lambda1}. This implies that $f_2$ is a density.  \hfill $\Box$

\bigskip

\textbf{Proof of Lemma \ref{lemma:lambda_sign}:}
We  again use the notation in \eqref{not}.
Let $\overline \lambda $ be  the  solution of equation (\ref{eq:lambda}) and $f_2^*$ be the ``optimal'' density defined by \eqref{f1F2opt}, then
\begin{align*}
\eta_1 - \eta_2 &= \int y f_1(y) dy - \int y f_2^*(y) dy = \int y f_1(y) \left\{ 1 - \frac{1}{1+\overline{\lambda}(y - \eta_2)} \right\} dy \\
&=\int y f_1(y) \frac{\overline\lambda (y - \eta_2)}{1+\overline\lambda (y - \eta_2)} dy = \overline\lambda \int y (y - \eta_2) f_2(y) dy  \\
&= \overline\lambda \int (y-\eta_2)^2 f_2(y) dy + \int (y\eta_2 - \eta_2^2) f_2(y) dy= \overline{\lambda} v_2^2,
\end{align*}
where $v_2^2 >0$ is the variance of $f_2(y)$,
and the last equality follows from the fact that
$$
\int y f_2(y) dy = \eta_2 .
$$
\hfill\ensuremath{\square}

\bigskip

\textbf{Proof of Theorem \ref{thm1}:}
Again we only prove part (a). Part (b) follows by similar arguments. 
Roughly speaking, the equivalence theorem is a consequence of the equivalence theorem for KL-optimal designs using the specific density for the second model $$f_2^*(y,x,\theta_2) = \frac{f_1(y,x,{ \overline\theta_1})}{1+\lambda(y-\eta_2(x,\theta_2))}$$ in the KL-criterion. More specifically, the criterion $K_{(a)}(\xi,{\overline\theta_1})$ is concave and to obtain the equivalence theorem, we calculate the directional derivative as follows:
\begin{align*}
\frac{\partial K_{(a)}([1-\alpha] \xi + \alpha \xi_{x^*},{\overline\theta_1})}{\partial \alpha} = \frac{\partial
\inf_{\theta_{2}\in \Theta_2}
\int_{\mathcal{X}} I_{1,2} (x, f_{1},f_{2} ,  {\overline{\theta}_{1}}, \theta_{2}) ([1-\alpha] \xi + \alpha \xi_{x^*}) (dx)}{\partial \alpha},
\end{align*}
where $\xi_{x^*}$ is the  Dirac measure at $x^*$. By theorem in \cite{Pshenichnyi1971} on page 75, and  the assumptions in our theorem, we have
\begin{align*}
\frac{\partial K_{(a)}([1-\alpha] \xi + \alpha \xi_{x^*},{\overline\theta_1})}{\partial \alpha} &=
\int_{\mathcal{X}} I_{1,2} (x, f_{1},f_{2} ,  {\overline{\theta}_{1}}, \widehat\theta_{2}) \xi_{x^*}(dx) - \inf_{\theta_{2}\in \Theta_2} \int_{\mathcal{X}} I_{1,2} (x, f_{1},f_{2} ,  {\overline{\theta}_{1}}, \theta_{2}) \xi (dx)\\
&=  I_{1,2} (x^*, f_{1},f_{2} ,  {\overline{\theta}_{1}}, \widehat\theta_{2}) - K_{(a)}(\xi,\overline\theta_1),
\end{align*}
where
\begin{align*}
\widehat\theta_2 = \arginf_{\theta_{2}\in \Theta_2} \int_{\mathcal{X}} I_{1,2} (x, f_{1},f_{2} ,  {\overline{\theta}_{1}}, \theta_{2}) \xi (dx).
\end{align*}
The design $\xi^*$ is optimal if and only if the inequality  
$$\frac{\partial K_{(a)}([1-\alpha] \xi^* + \alpha \xi_{x},{\overline\theta_1})}{\partial \alpha} \leq 0.$$
holds for every $x \in \mathcal{X}$, which means that we can not improve the concave functional $K_{(a)}(\xi^*,{\overline\theta_1})$ by moving in any direction from $\xi^*$.
\hfill\ensuremath{\square}

\bigskip

\textbf{Proof of Theorem \ref{thm4}:}
With the notations~\eqref{not} consider the function
\begin{align*}
F(\eta_1,\eta_2,\lambda)
&= \int_{-a+\eta_1}^{a+\eta_1} \log\left\{ 1 + \lambda (y - \eta_2) \right\} f_1(y) dy  \\
&= \int_{-a}^a \log\left\{ 1 + \lambda (t - \Delta_{\eta}) \right\} g(t) dt =: F(\Delta_{\eta},\lambda)~,
\end{align*}
where we have used  the representation \eqref{symass}  and the notation $\Delta_{\eta} = \eta_2 - \eta_1$.
Similarly, we obtain for the left-hand side of \eqref{eq:lambda} the representation
\begin{align*}
G(\eta_1,\eta_2,\lambda) = \int_{-a+\eta_1}^{a+\eta_1} \frac{f_1(y)}{1+\lambda (y - \eta_2)} dy = \int_{-a}^a \frac{g(t)}{1+\lambda (t - \Delta_{\eta})} dt.
\end{align*}
We now define the function
\begin{align*}
F(\Delta_{\eta}) = F(\Delta_{\eta},\overline{\lambda}(\Delta_{\eta})),
\end{align*}
where $\overline{\lambda}(\Delta_{\eta})$ is a   non-zero solution of  the equation
$G(\eta_1,\eta_2,\lambda) = 1$, which exists for all $x \in \mathcal{X}$ and for all $\theta_2 \in \Theta_2$ by the assumption of the theorem. Note that the function
\begin{align*}
h(t,\Delta_{\eta}) = \frac{g(t)}{1+\overline{\lambda}(\Delta_{\eta}) (t - \Delta_{\eta})}
\end{align*}
 is a density function with mean $\Delta_{\eta}$, because $g(t)$ is symmetric on $[-a,a]$ with mean $0$. For  the derivative of $F(\Delta_{\eta})$ we therefore obtain
\begin{align*}
\frac{\partial F}{\partial \Delta_{\eta}}(\Delta_{\eta}) &= \frac{\partial}{\partial \Delta_{\eta}} \int_{-a}^{a} \log\left\{ 1 + \overline{\lambda}(\Delta_{\eta}) (t - \Delta_{\eta}) \right\} g(t) dt \\
&= \int_{-a}^{a} \left[ \frac{\partial \overline{\lambda}}{\partial \Delta_{\eta}}(\Delta_{\eta}) (t - \Delta_{\eta}) - \overline \lambda(\Delta_{\eta}) \right] \frac{g(t)}{1+\overline{\lambda}(\Delta_{\eta})(t-\Delta_{\eta})} dt \\
&= \frac{\partial \overline{\lambda}}{\partial \Delta_{\eta}}(\Delta_{\eta})
\int_{-a}^{a} (t - \Delta_{\eta}) h(t,\Delta_{\eta}) dt
 - \overline{\lambda}(\Delta_{\eta}) \int_{-a}^{a} h(t,\Delta_{\eta}) dt = - \overline{\lambda}(\Delta_{\eta}).
\end{align*}
Here the first integral vanishes because
$h$  { has mean }  $\Delta_{\eta} $ and the second integral is equal to $1$ because $h$  is  a density.
From Lemma~\ref{lemma:lambda_sign}, we have the following implications:
If
$$ \eta_1 > \eta_2 ,  \text{ then ~} \Delta_{\eta} < 0 \Rightarrow \overline{\lambda}(\Delta_{\eta}) > 0 \Rightarrow F(\Delta_{\eta})
\text{ is decreasing, }
$$
and if
$$ \eta_1 < \eta_2,  \text{ then ~}  \Delta_{\eta} > 0 \Rightarrow \overline{\lambda}(\Delta_{\eta}) < 0 \Rightarrow F(\Delta_{\eta})
 \text{ is increasing. }
 $$
We also note that  the symmetry of $g$  implies  the symmetry of  $F$, i.e. $F(\Delta_{\eta}) = F(-\Delta_{\eta})$.
Let $\xi^*_T = \left\{ x^*_1,\dots,x^*_k; \omega^*_1,\dots,\omega^*_k \right\}$ be a $T$-optimal discriminating design maximizing~\eqref{topt} and define
\begin{align*}
\theta^*_2 = \arginf_{\theta_2} \int  [\eta_1(x,\overline\theta_1) - \eta_2(x,\theta_2)]^2 \xi^*_T(dx).
\end{align*}
By  the equivalence theorem for $T$-optimal designs  \citep{dettit2009}, it follows that for all  $x \in {\cal X} $, we have
\begin{align*}
\; |\eta_1(x,\overline\theta_1) - \eta_2(x,\theta^*_2)| \leq \epsilon = |\eta_1(x^*_1,\overline\theta_1) - \eta_2(x^*_1,\theta^*_2)| = \dots = |\eta_1(x^*_n,\overline\theta_1) - \eta_2(x^*_n,\theta^*_2)|.
\end{align*}
These arguments and the representation \eqref{f1F2opt} for the ``optimal'' density
$f_2^*$  yield for all $ x \in {\cal X}$
\begin{align*}
 I_{1,2}(x,f_1,f_2^*,\theta_1,\theta_2^*) &= F(\eta_1(x,\overline\theta_1) - \eta_2(x,\theta_2^*))  \\
& = F(|\eta_1(x,\overline\theta_1) - \eta_2(x,\theta_2^*)|) \leq F(\epsilon)   = \sum_{i = 1}^k \omega_i^* F(\epsilon)  \\
&= \sum_{i = 1}^{k} \omega_i^* F(\eta_1(x_i^*,\overline\theta_1) - \eta_2(x_i^*,\theta_2^*)) =
\int I_{1,2}(x,f_1,f_2^*,\theta_1,\theta_2^*)  d \xi^* (x) .
\end{align*}
Here the second equality follows from the symmetry of $F$ and the inequality is a consequence of the monotonicity of $F$.
By Theorem \ref{thm1} this  means that  the design $\xi^*$ is a  semi-parametric optimal design  for discriminating between
the model $f_1(y,x, \overline\theta_1)$ and the class $\mathcal{F}_2$.
\hfill\ensuremath{\square}

\bigskip

\textbf{Proof of Theorem \ref{thm5}:} 
We will show that the criterion~\eqref{f2F1symp} is equivalent to the criterion~\eqref{topthet} when $f_2(y,x,\theta_2)$ is a normal density. For simplicity, let
$
\eta_1 = \eta_1(x,\overline\theta_1)$, let  $\eta_2 = \eta_2(x,\theta_2)$ and let $ v_2 = v_2(x,\theta_2).
$
It follows from part (b) of Theorem \ref{prop1} that
\begin{align*}
f^*_1(y,x,\overline\theta_1) \propto f_2(y) \exp(-\lambda y) &= \frac{1}{\sqrt{2\pi} v_2} \exp\Big [-\frac{\left\{y-\eta_2\right\}^2}{2 v_2^2}-\lambda y\Big] \\
&= \frac{1 }{\sqrt{2\pi} v_2} \exp\Big[-\frac{\left\{y-(\eta_2-v_2^2 \lambda)\right\}^2}{2 v_2^2}\Big] \exp\Big[-\eta_2\lambda + \frac{v_2^2 \lambda^2}{2} \Big].
\end{align*}
The condition that  $f^*_1(y,x,\overline\theta_1)$ is  a density with mean $\eta_1$ yields 
\begin{align*}
\lambda = \frac{\eta_2 - \eta_1}{v_2^2}, \; \mu^\prime = \exp\Big[ \eta_2 \lambda - \frac{v_2^2\lambda^2}{2} \Big],
\end{align*}
 which implies that $f^*_1(y,x,\overline\theta_1)$ is a normal density with mean $\eta_1$ and variance $v_2^2$.
Then the KL-divergence between $f^*_1(y,x,\overline\theta_1)$ and $f_2(y,x,\theta_2)$ is given by
$
{[\eta_1 - \eta_2]^2}/{v^2_2},
$
which yields the criterion \eqref{topthet} and completes the proof. 
\hfill\ensuremath{\square}

\bigskip
\bigskip
\medskip

{\bf Acknowledgements.} Parts of this work were done during a visit of the
second author at the Department of Mathematics, Ruhr-Universit\"at
Bochum, Germany.
The work of H. Dette and R. Guchenko was supported by the Deutsche
Forschungsgemeinschaft (SFB 823: Statistik nichtlinearer dynamischer Prozesse, Teilprojekt C2).
The research of H. Dette  and W.K. Wong reported in this publication was also partially supported by the National Institute of
General Medical Sciences of the National Institutes of Health under Award Number R01GM107639.
The content is solely the responsibility of the authors and does not necessarily
 represent the official views of the National
Institutes of Health.
 The work of V. Melas and R. Guchenko was also partially supported by St. Petersburg State University
 (project "Actual problems of design and analysis for regression models", 6.38.435.2015).

\bigskip
\setstretch{1.2}
\setlength{\bibsep}{1pt}
\begin{small}
 \bibliographystyle{apalike}
\itemsep=0.5pt
\bibliography{model}
\end{small}
\end{document}